\documentclass{article}
\pdfoutput=1
\usepackage{jheppub}

\newcommand{\roughly}[1]{\mathrel{\raise.3ex\hbox{$#1$\kern-0.85em\lower1ex\hbox{$\sim$}}}}

\newcommand{\lsim}{\roughly<}
\newcommand{\gsim}{\roughly>}

\newcommand{\abs}[1]{\left|{#1}\right|}
\newcommand{\eq}[2]{\begin{equation}\label{#1}#2 \end{equation}}

\newcommand{\calR}{{\cal R}}

\newcommand{\mpl}{M_{p}}

\def\cA{{\cal A}}
\def\cB{{\cal B}} 
\def\cC{{\cal C}}

\def\cF{{\cal F}}
\def\cG{{\cal G}}

\def\cI{{\cal I}}
\def\cJ{{\cal J}}

\def\cL{{\cal L}}
\def\cM{{\cal M}}
\def\cN{{\cal N}}
\def\cO{{\cal O}}

\def\cQ{{\cal Q}}
\def\cR{{\cal R}}
\def\cT{{\cal T}}

\def\cV{{\cal V}}

\def\m{{\mu}}

\def\T{\mathcal{T}}
\def\curv{\mathcal{R}}
\def\N{\mathcal{N}}
\def\V{\mathcal{V}}
\def\A{\mathcal{A}}
\def\M{\mathcal{M}}
\def\F{\mathcal{F}}

\def\nn{\nonumber}
\def\({\left(}
\def\){\right)}
\def\[{\left[}
\def\]{\right]}

\newbox\charbox
\newbox\slabox
\def\slsh#1{{      
        \setbox\charbox=\hbox{$#1$}
        \setbox\slabox=\hbox{$/$}
        \dimen\charbox=\ht\slabox
        \advance\dimen\charbox by -\dp\slabox
        \advance\dimen\charbox by -\ht\charbox
        \advance\dimen\charbox by \dp\charbox
        \divide\dimen\charbox by 2
        \raise-\dimen\charbox\hbox to \wd\charbox{\hss/\hss}
        \llap{$#1$}
}}

\def\exd{{\hbox{d}}}

\def\nn{\nonumber}

\def\bea{\begin{eqnarray}}
\def\eea{\end{eqnarray}}
\def\be{\begin{equation}}
\def\ee{\end{equation}}

\def\ssM{{\scriptscriptstyle M}}
\def\ssN{{\scriptscriptstyle N}}

\def\EA{{\scriptscriptstyle EA}}
\def\EH{{\scriptscriptstyle EH}}

\def\pref#1{(\ref{#1})}

\title{Magnon Inflation: Slow Roll with Steep Potentials}

\author[a]{Peter Adshead,}
\author[b]{Diego Blas,}
\author[c,d]{C.P.~Burgess,}
\author[c,d]{Peter Hayman}
\author[e]{and Subodh P.~Patil}
\affiliation[a]{Department of Physics, University of Illinois at Urbana-Champaign, Urbana, IL 61801, USA }
\affiliation[b]{Theoretical Physics Department, CERN, CH-1211 Geneva 23, Switzerland}
\affiliation[c]{Physics \& Astronomy, McMaster University, Hamilton, ON, Canada, L8S 4M1}
\affiliation[d]{Perimeter Institute for Theoretical Physics, Waterloo, Ontario N2L 2Y5, Canada }
\affiliation[e]{Department of Theoretical Physics, University of Geneva, 24 Quai Ansermet, Geneva, CH-1211, Switzerland}

\emailAdd{adshead@illinois.edu, diego.blas@cern.ch, cburgess@perimeterinstitute.ca, haymanpf@mcmaster.ca, subodh.patil@unige.ch}

\preprint{Preprint: CERN-TH-2016-067}

\date{\today}

\abstract { We find multi-scalar effective field theories (EFTs) that can achieve a slow inflationary roll despite having a scalar potential that does not satisfy $\cG^{ab} \partial_a V \partial_b V \ll V^2/\mpl^2$ (where $\cG_{ab}$ is the target-space metric). They evade the usual slow-roll conditions on $V$ because their kinetic energies are dominated by single-derivative terms rather than the usual two-derivative terms. Single derivatives dominate during slow roll and so do not require a breakdown of the usual derivative expansion that underpins calculational control in much of cosmology. The presence of such terms requires some sort of UV Lorentz-symmetry breaking during inflation (besides the usual cosmological breaking). Chromo-natural inflation provides one particular example of a UV theory that can generate the multi-field single-derivative terms we consider, and we argue that the EFT we find indeed captures the slow-roll conditions for its background evolution. We also show that our EFT can be understood as a multi-field generalization of the single-field Cuscuton models. The multi-field case introduces a new feature, however: the scalar kinetic terms define a target-space 2-form, $\cF_{ab}$, whose antisymmetry gives new ways for slow roll to be achieved.
}

\begin{document}
\maketitle
\section{Introduction}

Primordial fluctuations provide a rare observational window into the high-energy physics of the pre-nucleosynthesis universe. Remarkably, the observed properties of these fluctuations are consistent with 
vacuum fluctuations stretched out to very large scales  by the accelerated expansion of spacetime \cite{Starobinsky:1979ty, Starobinsky:1980te,Lukash:1980iv,Press:1980zz, Mukhanov:1981xt,Guth:1982ec,Hawking:1982cz}. Much effort has been invested in determining 
the origins of both the fluctuations and the accelerated expansion, with inflationary models \cite{Starobinsky:1980te, Guth:1980zm,Linde:1981mu,Albrecht:1982wi} emerging as the simplest framework within which both are understood within a controlled semiclassical approach. 

Simple phenomenological models of inflation are easy to write down \cite{Linde:1983gd}, typically relying on slowly rolling scalar fields. These models require slow roll in order to exploit a proximity to the exponentially expanding de Sitter geometry that is obtained when gravity is dominated by a static scalar potential energy, $T_{\mu\nu} = \cV_0 \, g_{\mu\nu}$. The scalar motion (and its gravitational response) can be analyzed using a derivative expansion, leading to generic inflaton Lagrangians of the form\footnote{Because we use dimensionless fields a squared-mass scale, $F^2$, is extracted from the scalar kinetic term in order to allow the target-space metric, $\cG_{ab}$, to be dimensionless. We adopt MTW curvature conventions in what follows.}
\be \label{2derivs}
 \cL = - \sqrt{-g} \left\{ \cV(\phi) + \frac12 \, g^{\mu\nu} \Bigl[F^2 \,\cG_{ab} (\phi) \, \partial_\mu \phi^a \partial_\nu \phi^b - \mpl^2 \, R_{\mu\nu} \Bigr] + \cdots \right\} \,,
\ee
where $R_{\mu\nu}$ is the metric's Ricci tensor and $\cV(\phi)$ and $\cG_{ab}(\phi)$ are two functions whose specification defines the precise model. 

The ellipses in \pref{2derivs} denote terms with more than two derivatives, and these are normally negligible to first approximation precisely because the assumed scalar motion is slow. They can in principle play a role at higher order, with terms like $(\partial \phi)^4$ introducing small differences between the relevant speed of sound and that of light, $\delta c_s^2 = 1 - c_s^2$, but the whole framework of expanding in derivatives becomes suspicious to the extent that these corrections are large.\footnote{Higher derivative terms can play a more significant role during inflation \cite{ArmendarizPicon:1999rj}, but semiclassical calculations using such terms are only under control \cite{Burgess:2009ea, Donoghue:2012zc, Goldberger:2007hy,Burgess:2003jk} to the extent that the underlying approximation is no longer a simple derivative expansion. DBI Inflation \cite{Alishahiha:2004eh, Chen:2008hz, deRham:2010eu} is the poster child for such models, where an implicit (non-linearly realized) Lorentz symmetry protects the expansion (see also \cite{Achucarro:2012yr} for an example with an emergent shift symmetry).}

Despite the simplicity of derivative expansions, such models prove to be notoriously difficult to embed plausibly into a sensible physical framework at higher energies (see, however, \cite{Burgess:2011fa, Cicoli:2011zz,Baumann:2009ds,Baumann:2009ni,McAllister:2007bg,Burgess:2007pz,Quevedo:2002xw}). A major reason for this is the fairly generic slow-roll requirement for very shallow scalar potentials. This is usually quantified by asking sufficiently small values for the slow-roll parameters \cite{Liddle:1992wi},
\be \label{vepst}
 \varepsilon_{\rm st} = \frac{\mpl^2}{2F^2} \; \cG^{ab}  \left( \frac{ \cV_{,a} \cV_{,b}}{\cV^2} \right) \lsim 10^{-2} \,,
\ee
(plus a similar condition on second derivatives of $\cV$ to ensure that inflation lasts sufficiently long) where $\cV_{,a}$ denotes $\partial_a \cV$ while $\cG^{ab}$ is the inverse target-space metric, defined by $\cG^{ab} \cG_{bc} = \delta^a_c$. These conditions are difficult from the point of view of a UV completion both because it is typically hard to arrange sufficiently small derivatives --- e.g. $\cV_{,a}/\cV$ in \pref{vepst} --- and to get\footnote{This is most crisply stated within string theory, for which axion-like Goldstone bosons -- such as generate the trigonometric potentials of `natural inflation' models \cite{Freese:1990rb} -- typically satisfy $F \lsim M_s \ll \mpl$, while scale-breaking Goldstone-boson inflatons \cite{Burgess:2014tja, Csaki:2014bua} -- such as arise for extra-dimensional moduli \cite{Lust:2013kt,  Blumenhagen:2012ue, Cicoli:2011ct, Cicoli:2008gp, Bond:2006nc, Conlon:2005jm, Burgess:2001vr} --- have $F \lsim \mpl$.} $F \gsim \mpl$ (see, however, \cite{Kim:2004rp, Bachlechner:2014gfa, Choi:2014rja, Long:2014dta, Burgess:2014oma, Ben-Dayan:2014lca}).

The purpose of this paper is to try to evade these obstructions by describing a new class of inflationary models that arise within a systematic derivative expansion, but for which inflationary slow-roll does not require either shallow potentials or trans-Planckian values for $F$. We do so by supplementing the Lagrangian of \pref{2derivs} --- {i.e.} $\cL \to \cL + \Delta \cL$ --- with terms with {\em fewer} than two derivatives. The new terms can easily be arranged to dominate the two-derivative terms of \pref{2derivs} during slow roll without having unnaturally large coefficients. Single derivative terms in $\cL$ are nominally excluded by Lorentz invariance, and so we are forced to work within a framework wherein Lorentz symmetry is broken in the UV. We mostly focus on the case where this breaking is characterized by a timelike 4-vector order parameter, $U^\mu$, whose expectation chooses a preferred frame. 
Other symmetry breaking patterns include Chromo-natural inflation \cite{Adshead:2012kp} (see also \cite{Maleknejad:2011jw}), where a UV completion is known  (see also  \cite{Dimastrogiovanni:2012st} for an effective description). 

The new interactions we explore are given by the term
\be \label{1derivdef}
 \Delta \cL = - \sqrt{-g} \; \cA_a(\phi) \, U^\mu \partial_\mu \phi^a \,,
\ee
where $\cA_a(\phi)$ is a new set of coefficient functions that must be specified and $U^\mu$ is a time-like unit vector.  When relevant, we also consider the additional interactions that arise at the two-derivative level due to the presence of the new field $U^\mu$, but our main focus is when the term of \pref{1derivdef} dominates. For dimensionless fields $\cA_a$ has dimensions $\mu^3$ for a UV scale $\mu$, whose value may or may not be related to $F$, or the scale $M$ of Lorentz breaking (more on this in sec.~\ref{sec:fulltheory}). Because the Lagrangian \pref{1derivdef}  --- together with the spatial-derivative $(\nabla \phi)^2$ terms --- describes spin waves within ferromagnets\footnote{Because a ferromagnet breaks time-reversal invariance, ferromagnetic spin waves have low-energy dispersion relations $\omega \propto k^2$, unlike the more familiar $\omega^2 \propto k^2$ dispersion of magnons in antiferromagnets.} \cite{Burgess:1998ku, Watanabe:2012hr},  we call this class of models {\em Magnon Inflation}. 

We compute how slowly rolling scalar fields evolve when governed by $\cL + \Delta \cL$ and identify the circumstances where the two-derivative terms are dominated by the potential and the one-derivative term of $\Delta \cL$. When $\mu \sim F \sim M$ this happens for frequencies, $\Gamma$, satisfying $\Gamma \ll M$. Computing the gravitational response gives a different dependence of the slow-roll parameters on the scalar potential, for instance with \pref{vepst} replaced by 
\be \label{1derivnew}
 \varepsilon := - \frac{\dot H}{H^2}  = \frac32 \left( \frac{ \widetilde \cF^{ab} \cA_a   \cV_{,b} }{\cV } \right) \,,
\ee
with
\be
\label{eq:Fab}
\cF_{ab} := \partial_a \cA_b - \partial_b \cA_a, \quad \quad \widetilde \cF^{ab} \cF_{bc} = \delta^a_c,
\ee
assumed to be non-degenerate. This depends very differently on the scales of the Lagrangian and so imposes qualitatively different slow-roll conditions on $\cV$. In particular, the antisymmetry of $\widetilde\cF^{ab}$ implies $\varepsilon$ vanishes identically whenever the target-space vectors $\cA_a$ and $\cV_{,a}$ are parallel to one another {\em regardless} of how steep the scalar potential is. Notice that at least two scalar fields are required to have nonzero $\cF_{ab}$ and so for eq.\ \eqref{1derivnew} to apply.

It is noteworthy that the sign of eq.\  \pref{1derivnew} need not be positive when nonzero, in contrast to eq.\  \pref{vepst}. Since $\varepsilon \propto p + \varrho$, this means that, for some parameters, magnon models can violate the Null Energy Condition (NEC). Although this need not imply instability in general \cite{Rubakov:2014jja}, whether it does or not must be checked in any particular instance. We examine stability for these models and argue that they can be stable for both signs of $\varepsilon$. The models with vanishing $\varepsilon$ are typically marginally stable if only the single-derivative terms are included. An assessment of the stability of slow evolution for these models requires including the leading higher-derivative terms, and we evaluate the combination of two-derivative couplings that controls the sign of $\varepsilon$ in simple models. 

For nonzero $\varepsilon$ the stability analysis can be done by explicitly integrating out all constraints, leading to a classically equivalent theory involving a multiple-field but single-clock (and finite speed of sound) generalization of the Cuscuton models considered in \cite{Afshordi:2006ad, Afshordi:2007yx, Afshordi:2009tt}. As special cases, this equivalent classical reformulation also contains other standard inflationary models, such as canonical and derivatively coupled $P(X)$ models \cite{ArmendarizPicon:1999rj}. 

Magnon inflation also superficially resembles Ghost Inflation \cite{ArkaniHamed:2003uz,Senatore:2004rj,Ivanov:2014yla}, Inflaton-Aether models \cite{Donnelly:2010cr,Solomon:2013iza}, and $\Theta$CDM   \cite{Blas:2011en} inasmuch as these also include Lorentz-breaking interactions that are linear in time derivatives.  However, these other models usually involve only a single scalar field and as a result $\cF_{ab}$ vanishes.

\section{Lowest derivative action}
\label{sec:lowestderiv}

This section outlines the action, field equations and conserved quantities for the system of interest, including only the leading derivative interactions. The main point is to show that the slow-roll parameters can vanish completely in some cases for this action, despite the Lagrangian including a potential subject to no steepness conditions. We return in later sections to how the dominant subleading corrections modify this picture.

\subsection{Action, scales and field equations}\label{sec:fulltheory}

We examine the mutual interactions of a collection of scalars, $\phi^a$, and the metric, $g_{\mu\nu}$. These interactions happen at energies well below a scale $M$ characterizing the breaking of Lorentz invariance. We focus in the scenario where this happens through an order parameter, $U^\mu$, which transforms as a time-like contravariant 4-vector and whose magnitude is heavy enough to be frozen at low energies (so $g_{\mu\nu} U^\mu U^\nu = -1$). It is useful  to enforce this condition through the term in the action
\be
 S_{\rm \xi} = - \int \exd^4x \sqrt{-g} \; \xi \, \Bigl( g_{\mu\nu} U^\mu U^\nu + 1 \Bigr) \,,
\ee
where $\xi$ is a Lagrange-multiplier field.\footnote{The special case where $U^\mu$ is hyper-surface orthogonal is the so-called khronometric
case, though we do not here restrict ourselves to this case (see \cite{Blas:2010hb,Jacobson:2010mx} for a discussion relating the more general Einstein-aether and khronometric preferred frame scenarios).}

The most relevant operators are  given by all possible interactions with the lowest number of derivatives in each sector:  i.e.\  $S = S_\EH + S_\ssM$ where $S_\EH$ is the standard Einstein-Hilbert action and
\be
\label{action1}
 S_\ssM = - \int \exd^4x \sqrt{-g} \; \Bigl[ \cV(\phi) + \cA_a(\phi) \, U^\mu \partial_\mu \phi^a + \cB(\phi) \nabla_\mu U^\mu  + \cdots \Bigr] \,.
\ee
Here the ellipses involve terms with more than two derivatives and all of the independent fields, $\phi^a$, $U_\mu$ and $g_{\mu\nu}$ are to be varied. We assume that whatever the UV Lorentz-violating physics is, it does not also generate an $\cO(M^4)$ contribution to the scalar potential, $\cV$ (i.e.\  we do not solve the cosmological-constant problem).

Notice that if we make the substitutions $\delta \cA_a = \partial_a \Omega$ and $\delta \cB = \Omega$ for any scalar target-space function, $\Omega = \Omega(\phi)$, the change in the action becomes
\be
 \delta S_\ssM = - \int \exd^4 x  \sqrt{-g} \; \Bigl[ U^\mu \partial_\mu \phi^a \,  \partial_a \Omega(\phi) +\Omega(\phi) \nabla_\mu U^\mu \Bigr] 
 = -  \int \exd^4 x \partial_\mu \Bigl[ \sqrt{-g} \; U^\mu \Omega (\phi) \Bigr] \,,
\ee
which reveals this to be a symmetry of the classical equations, up to surface terms. When boundary effects are not important it is useful to use this symmetry to choose the gauge $\cB = 0$, as we now do.\footnote{For applications where $\nabla \cdot U = 0$ (such as for flat space) this argument also reveals $\cA_a$ to be a gauge potential on the target space (up to boundary terms), with physical quantities only depending on `gauge-invariant' combinations like $\cF_{ab} = \partial_a \cA_b - \partial_b \cA_a$ \cite{Burgess:1998ku}.}
Notice that the possible distinction between $\cA_a$ and $\cB$  disappears in the case where there is only one scalar field because we can always choose $\cA = - \cB'$. In particular, $\cF_{ab}=0$ in the single-field case.

\subsubsection*{Scales}

\begin{figure}[h]
\begin{center}
\includegraphics[width=0.7\textwidth]{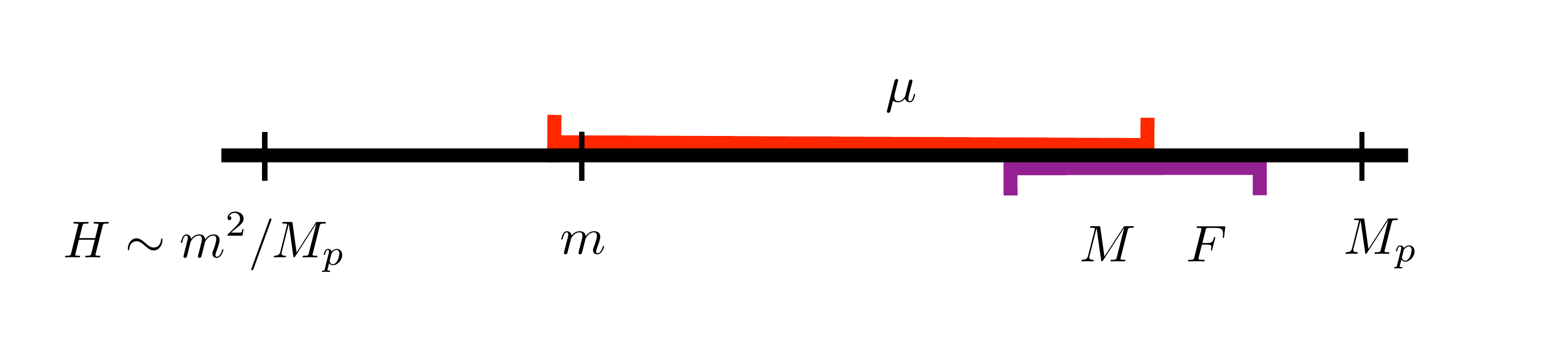}
\caption{Scales relevant for magnon inflation. The are three UV scales: $M\sim F$  and $\mpl$ and two infrared scales $m$ and $H$.
The scale $m$ is assumed to be protected by some symmetry. The scale $\m$ can lie in a wide interval without compromising
naturalness (see the main text)}
\label{fig:scales}
\end{center}
\end{figure}

Before exploring the field equations following from the previous action we first pause to discuss the scales implicit in the problem, since these --- together with the derivative expansion that underlies the entire formalism --- define the domain of validity of any such analysis. For the purposes of doing so it is convenient to work with dimensionless fields, $\phi^a$ and $U^\mu$. The action defines the following important energy scales (summarized in figure \ref{fig:scales}):
\begin{itemize}
\item {\em Gravitational response:} We denote the coefficient of the Einstein-Hilbert action by $\mpl^2$, though once Lorentz-breaking fields like $U^\mu$ are present, the coupling $G$ defined by $8 \pi G = \mpl^{-2}$ need not be the precise physical Newton constant, $G_\ssN$, as measured, say, in the solar system or in cosmology \cite{Jacobson:2008aj,Blas:2014aca}. Our interest in what follows is mainly situations for which $G$ is of the same order of magnitude as $G_\ssN$, and so for which $\mpl$ has its traditional order of magnitude.
\item {\em Scalar kinetic energy:} At high enough energies two-derivative scalar interactions are no longer negligible relative to the single-derivative terms considered up to this point. For dimensionless scalars the kinetic terms are multiplied by a scale, which we denote by $F^2$. The scale $F$ can be, but need not be, of order $\mpl$, depending on the origin of the scalars. For instance, for would-be goldstone bosons $F$ is of order the size of the vev that spontaneously breaks the corresponding approximate symmetry.
\item {\em $U^\mu$ kinetic energy:} Kinetic terms for the field $U^\mu$ are also possible once two-derivative interactions are considered and are also accompanied by a scale, which we denote by $M^2$. Just as for the scalars, the scale $M$ is likely the scale at which the
dynamics of the order parameter $U^\m$ is inevitably modified. 
\item {\em Time-reversal breaking:} We denote the energy scale set by the single-derivative terms by $\mu$. For dimensionless fields,
this means that  $\cA_a \sim \mu^3$ in order of magnitude. Because these terms are the lowest-dimension interactions that break both time-reversal and Lorentz invariance, it can be natural for $\mu$ to be much smaller than other scales like $M$ and $\mpl$.
\item {\em Scalar potential energy and $H$:} For dimensionless fields we denote the generic energy scale set by the potential to be $m$, so the potential is $\cV = m^4 v(\phi)$ where $v(\phi)$ is a dimensionless function. It is possible for this scale to be much smaller than the previous scales if the scalars enjoy an approximate shift symmetry, as indeed would be the case for pseudo-Goldstone bosons. For inflationary applications, we assume that the Hubble scale during inflation is $H \sim m^2/\mpl$, and so we take $m \ll \mpl$ and so also $H \ll m$. 
\end{itemize}

In the applications envisioned here we regard all of the scales\footnote{The UV Lorentz-breaking scale $M$ is subject to strong observational constraints should the field $U^\mu$ survive into the present-day universe \cite{Blas:2014aca}. We remain agnostic to this possibility and here focus only on its influence on inflationary predictions.} ---  i.e.\ $F$, $M$ and $\mu$ --- as being bounded above by the Planck scale, $M,F,\mu \lsim \mpl$, while at the same time being much larger than the natural scale, $\Gamma \sim \dot \phi/\phi$, of the time-dependence of any background evolution. In making estimates below for simplicity we often assume all UV scales to be similar in size: $M \sim F \sim \mu \lsim \mpl$. 

For order-unity fields --- i.e.\ $\phi , U^\mu \sim \cO(1)$ --- the action's two derivative terms are order $F^2 \dot\phi^2 \sim F^2 \Gamma^2$ and $M^2 (\nabla U)^2 \sim M^2 \Gamma^2$ in size, so their neglect relative to the one-derivative term --- whose size is $\cA_a \dot \phi^a \sim \Gamma \mu^3$ --- requires $\Gamma$ to satisfy
\be \label{Ebounds}
  \Gamma  \ll \hbox{min} \left( \frac{\mu^3}{F^2} \; , \; \frac{\mu^3}{M^2} \right) \,,
\ee
which becomes $\Gamma \ll M$ when $F \sim \mu \sim M$. For cosmological applications this must hold in particular for $\Gamma \sim H \sim m^2/\mpl$, where, as above, $\cV \sim m^4$ sets the scale of the scalar potential. Figure \ref{fig:scales} indicates how the UV scales are related and shows that \pref{Ebounds} can easily be accommodated if $m$ is much smaller than the other scales in the problem.

\subsubsection*{Field equations}

The equations of motion for the fields $\xi$ and $U^\mu$ are algebraic,
\be \label{zetaeom}
 g_{\mu\nu} U^\mu U^\nu = -1 \,,
\ee
and
\be \label{Ueom}
 \cA_a(\phi)  \, \partial_\mu \phi^a + 2\xi \, U_\mu   = 0 \,.
\ee
Contracting \pref{Ueom} with $U^\mu$ and using \pref{zetaeom} then gives an expression for $\xi$: 
\be \label{zetaeomsoln}
 2\xi =  \cA_a(\phi) \,  U^\mu \partial_\mu \phi^a \,.
\ee
Notice that in the special case $\cA_a = \partial_a \Omega$ (and for nonzero $\xi$) eq.\ \pref{Ueom} implies $U_\mu$ is proportional to a gradient, $U_\mu = - (2\xi)^{-1} \, \partial_\mu \Omega$ and so is hypersurface-orthogonal, with the level-surfaces of $\Omega$ providing a natural notion of time.  

Our interest is in backgrounds  for which the scalars depend only on a time coordinate, $\phi^a = \phi^a(t)$, and thus define a `cosmic' frame (with 4-velocity $u^\mu$).
In this case, equation \eqref{Ueom} implies the aligned solution  $U^\m=u^\m$ (whenever $\xi\neq 0$).  This corresponds to a homogeneous and isotropic situation for which it is natural to look for metrics of the standard FRW form,
\be
\label{eq:FRW}
 \exd s^2 = - \exd t^2 + a^2(t) \, g_{ij}(x) \exd x^i \, \exd x^j \,.
\ee
This yields
\be
\label{eq:DU}
 \nabla_\mu U^\nu = \nabla_\mu u^\nu = H \left( \delta_\mu^\nu + U_\mu U^\nu \right)\,,
\ee
and so in particular $\nabla \cdot U = \nabla \cdot u = 3H$ where, as usual, $H = \dot a/a$. Of course homogeneity and isotropy themselves imply $U^\mu \propto u^\mu$ even if $\xi = 0$.\footnote{It has been argued more generally that a $U^\mu$ not initially aligned with $u^\mu$ often evolves to align with $u^\mu$ at later times once the two-derivative terms in the action are also included \cite{Kanno:2006ty,Carruthers:2010ii}.}

The scalar field equation is (recalling \eqref{eq:Fab})
\be \label{Seom}
 -\partial_a \cV - \cF_{ab} \, U^\mu \partial_\mu \phi^b + \cA_a \, \nabla \cdot U = 0 \,.
\ee
 Notice that if the target space is two-dimensional, then $\cF_{ab} \propto \, \epsilon_{ab}$ where $\epsilon_{ab}$ is the target-space volume form. In the one-dimensional case, the term involving derivatives of $\phi^a$ cancels which
 obscures the analysis of the propagating degrees of freedom at scales dominated by the single-derivative term (cf. section \ref{sec:cosmicfluc}).  

If the inverse of $\cF_{ab}$ exists, then \pref{Seom} implies
\be \label{Seom1}
 \dot \phi^a := U^\mu \partial_\mu \phi^a = \widetilde \cF^{ab} \Bigl[ \cA_b \, \nabla \cdot U - \partial_b \cV \Bigr] \,,
\ee
and so the antisymmetry of $\widetilde \cF^{ab}$ means that $\dot\phi^a$ is orthogonal (in the tangent to the target-space manifold) to $\partial_b \cV - \cA_b \, \nabla \cdot U$. Combined with \pref{zetaeomsoln} this gives
\be \label{zeta0eq}
 2 \xi =  \cA_a \,  U^\mu \partial_\mu \phi^a = - \widetilde \cF^{ab} \cA_a  \, \partial_b \cV \,.
\ee

The Einstein equation relates $G^{\mu\nu} = R^{\mu\nu} - \frac12 \, R \, g^{\mu\nu}$ to the stress-energy, $T^{\mu\nu}$, which is given by
\bea \label{zerothT}
 T^{\mu\nu} = \frac{2}{\sqrt{-g}} \left( \frac{\delta S_m}{\delta g_{\mu\nu}} \right)
 &=& - g^{\mu\nu} \left( \cV + \cA_a \, U^\lambda \partial_\lambda \phi^a \right) - 2\xi \, U^\mu U^\nu \\
 &=& - g^{\mu\nu} \; \cV - \cA_a \, \dot \phi^a \Bigl(  g^{\mu\nu} + U^\mu U^\nu \Bigr) \nn\\
 &=& - g^{\mu\nu}  \; \cV + \widetilde \cF^{ab}  \cA_a \, \partial_b \cV  \Bigl(  g^{\mu\nu} + U^\mu U^\nu \Bigr) \,, \nn
\eea
and so $U_\nu T^{\mu\nu} = - \varrho \, U^\mu$, where the energy density is given by
\be\label{eqn:energy}
 \varrho := U^\mu U^\nu T_{\mu\nu} = \cV \,.
\ee
Similarly the pressure is
\be
\label{pressure}
 p := N^\mu N^\nu T_{\mu\nu} = \widetilde \cF^{ab}  \cA_a \, \partial_b \cV  - \cV \,,
\ee
for any spacelike $N^\mu$ satisfying $N\cdot U = 0$ and $N \cdot N = 1$.

Notice that the condition $\widetilde \cF^{ab}  \cA_a \, \partial_b \cV  = 0$ is sufficient to ensure $p = - \varrho$, such as is true when $\cA_a$ is parallel in field space to $\partial_a \cV$. In this case $\varrho$ is constant and spacetime is de Sitter.\footnote{Notice that because $\xi$ also vanishes in this case no preferred time-slicing exists even if $\cA_a = \partial_a \Omega$.} 

\subsection{Slow-roll parameters}

Given the aligned configuration $U^\m=u^\m$, the Friedmann equation becomes
\be \label{Friedmann}
 3 \mpl^2 H^2 = \varrho = \cV \,.
\ee

We seek the slow-roll conditions to contrast with those for scalars with a regular kinetic term. The first slow-roll parameter is
\be \label{epsilon1stderiv}
 \varepsilon := - \frac{\dot H}{H^2} = \frac{\varrho + p}{2 \mpl^2 H^2} = \frac32 \left( \frac{ \widetilde \cF^{ab} \cA_a   \partial_b \cV }{\cV } \right) \,,
\ee
where the first line uses the Friedmann equation, eq.\  \pref{Friedmann}, and its rate of change together with stress-energy conservation, $\dot \varrho = - 3H(\varrho + p)$. Note that, using the equation of motion for $\phi^a$, eq.\ \eqref{Seom1}, this equation may also be expressed as
\eq{alt-eps}{\varepsilon = -\frac{\dot\varphi^c\cA_c}{2 H^2\mpl^2 }.}
In order of magnitude, writing $\cV \sim m^4 v(\phi)$ and $\cA \sim \mu^3 \alpha(\phi)$ --- and so also $\cF \sim \mu^3 \alpha' (\phi)$ --- eq.\ \eqref{epsilon1stderiv} implies $\varepsilon \sim (v'/v)/(\alpha'/\alpha)$. Since  both scales $\mu$ and $m$ dropped out,  asking $|\varepsilon| \ll 1$ generically demands $v'/v \ll \cO(1)$ (and the condition can be even weaker than this -- see below). The corrections to the previous
formula do not modify the slow-roll condition provided  $F^2/\mpl^2\ll 1$. Notice finally that $\varepsilon=0$ corresponds to $\xi=0$ (cf. eq.\ 
\eqref{zeta0eq}) for which the dynamics of $U^\mu$ is governed by the higher order operators (still, the aligned configuration can be a solution).

The expression at eq.\ \eqref{epsilon1stderiv} is to be compared with the usual slow-roll equation that follows from a two-derivative Lagrangian of the form of eq.\ 
\eqref{2derivs}. In this case slow roll would predict $3H \dot \phi^a \simeq - \cG^{ab} \partial_b \cV$ and $\frac12 \cG_{ab} \dot \phi^a \dot \phi^b \ll  \cV$ and so give the standard formula
\be
 \varepsilon_{\rm st} \simeq \frac32 \left( \frac{ \cG_{ab} \dot \phi^a \dot \phi^b}{\cV} \right) \simeq \frac{\mpl^2}{2} \left( \frac{\cG^{ab} \partial_a \cV \partial_b \cV}{\cV^2} \right)\sim \left(\frac{\mpl v'}{F v}\right)^2 \,,
\ee
instead of \pref{epsilon1stderiv}.  The requirement $\varepsilon_{\rm st} \ll 1$ asks the dimensionless function $v$ to satisfy $(v'/v)^2 \ll F^2/\mpl^2$. Given that most known systems give $F \lsim \mpl$ the conditions on $\cV$ required to ensure $|\varepsilon| \ll 1$ are generically weaker than those required to achieve $\varepsilon_{\rm st} \ll 1$. 

The conditions for small $\varepsilon$ can be even weaker than just asking $v'/v$ to be order unity. This is because the antisymmetry of $\cF_{ab}$ means that $\varepsilon$ can vanish identically even if $\partial_a \cV \ne 0$, such as if $\cA_a$ is parallel in field space to $\partial_a \cV$.  Notice also that because $\widetilde \cF^{ab}$ is antisymmetric eq.\ \pref{epsilon1stderiv} does not imply $\varepsilon$ must be nonnegative (or, equivalently, it allows $p + \varrho$ to be negative) and so the motion need not satisfy the NEC. We return below --- see section \ref{sec:stability} --- to whether or not this should give us pause.

When $\varepsilon$ does not vanish identically, the second slow-roll parameter can be evaluated using
\be 
 \eta := \frac{\dot \varepsilon}{H \varepsilon} = \dot \phi^a \, \frac{ \partial_a \varepsilon}{H \varepsilon} \,,
\ee
in which we can evaluate $\dot \phi^a$ using eq.\ \pref{Seom1} to get
\be \label{etaexp}
 \eta =  \left( \frac{ \partial_a \varepsilon}{H \varepsilon} \right) \widetilde \cF^{ab} \Bigl( 3H \cA_b - \partial_b \cV \Bigr) \,.
\ee
Here $\partial_a \varepsilon$ is evaluated by differentiating eq.\ \pref{epsilon1stderiv}. In order of magnitude this implies $\eta \sim y (\varepsilon'/\varepsilon)$ with $y \sim \alpha/\alpha'$ or $y \sim (m^4 / H \mu^3)(v'/\alpha') \sim (H \mpl^2 / \mu^3)(v'/\alpha')$, depending on which term dominates in the rightmost bracket of eq.\ \pref{etaexp}. The suppression of the latter term by $H$ ensures its contribution can be small if $\mu \lsim \mpl$, leading to a generic condition on $\alpha/\alpha'$. The antisymmetry of $\widetilde \cF^{ab}$ potentially allows even this condition to be avoided if $\partial_a \varepsilon$ is appropriately aligned relative to $3H \cA_b - \partial_b \cV$.

\subsection{An equivalent effective description} \label{sec:effdesc}

In this section we derive a classically equivalent reformulation of the above single-derivative model involving only scalar fields but with a more complicated kinetic sector. This reformulation is only possible when $\xi \ne 0$ and is obtained by integrating out the non-dynamical fields $U^\mu$ and $\xi$. 

We begin with the one-derivative magnon inflation action given earlier (repeated here for convenience)
\be
\label{act1}
 S_\ssM =  \int \exd^4\,x \sqrt{-g} \; \left\{ -\cV(\phi) - \cA_a(\phi) U^\mu \partial_\mu \phi^a - \xi \, \Bigl( g_{\mu\nu} U^\mu U^\nu + 1 \Bigr) \right\}.
\ee
If we take this action at face value (i.e. neglecting higher derivative contributions), the $U^{\mu}$ appears in the action as an auxiliary field and its functional integral is gaussian so we can straightforwardly integrate it out once and for all. The result is equivalent to evaluating \pref{act1} at the saddle point for $U^\mu$ found from its equation of motion, eq.\ \eqref{Ueom}
\eq{usol}{U_\mu = -\frac{1}{2\xi}\cA_a\partial_\mu\phi^a  \,,}
and substituting the above back into (\ref{act1}). (We see here why $\xi = 0$ must be avoided in this reformulation.) The result is
\be
\label{act22}
 S_\ssM = \int \exd^4\,x \sqrt{-g} \; \left\{ - \cV(\phi) + \frac{1}{4\xi}\cA_a\cA_b\partial_\mu \phi^a\partial^\mu\phi^b - \xi \right\} \,.
\ee

Similarly solving for $\xi$ using the saddle-point approximation (this time not an exact result, but perfectly adequate for the classical applications of interest), we find
\begin{align}\label{eqn:xisol}
\frac{1}{4\xi^2} \, \cA_a\cA_b\partial_\mu \phi^a\partial^\mu\phi^b +1 =0 
\quad \hbox{and so} \quad
\xi = \frac{\iota}{2}\sqrt{-\cA_a\cA_b\partial_\mu \phi^a\partial^\mu\phi^b} \,.
\end{align}
Here we use eq.\ \eqref{zetaeomsoln} to resolve the apparent sign ambiguity in taking the square root, with $\iota = {\rm sign}(\A_a \dot{\phi}^a) =  {\rm sign}(\A_a U^\mu \partial_\mu {\phi}^a)$. Substituting this into \pref{act22} then leads to
\eq{cusc}{ S_\ssM = -\int \exd^4x \sqrt{-g} \; \left\{  \cV(\phi) + \iota\sqrt{-\cA_a\cA_b\partial_\mu \phi^a\partial^\mu\phi^b} \right\} \,. }

Thus magnon inflation is classically equivalent to \pref{cusc}, which bears a superficial resemblance to the Cuscuton model \cite{Afshordi:2006ad, Afshordi:2007yx, Afshordi:2009tt} (see also \cite{Lim:2010yk}), though with multiple fields and with a dyadic (and so degenerate --- more about which below) target space metric: $\widetilde{\cG}_{a b}(\phi) = \cA_a\cA_b$. (we show in section \ref{sec:cosmicfluc}, however, that magnon inflation has perturbations that can propagate at finite $c_s$, unlike Cuscuton models.) Notice that the appearance of the square root in (\ref{cusc}) indicates the alternative formulation runs into trouble whenever $g^{\mu\nu} \cA_a \partial_\mu \phi^a \cA_b \partial_\mu \phi^b > 0$, which in our metric conventions corresponds to the vector $\cA_a \partial_\mu \phi^a$ becoming space-like. When $\cA_a \partial_\mu \phi^a$ is space-like, $U^\mu$ cannot be time-like (due to eq.\ \pref{usol}), which is required by the constraint $U^\mu U_\mu = -1$ enforced by the $\xi$ integration. 

To verify classical equivalence with the original formulation of the theory, we calculate the energy-momentum tensor
\eq{}{T_{\mu\nu} = -\frac{2}{\sqrt{-g}}\frac{\delta S}{\delta g^{\mu\nu}} = g_{\mu\nu}\cL - \frac{\iota\, \cA_a\cA_b\partial_\mu\phi^a\partial_\nu\phi^b}{\sqrt{-\cA_c\cA_d\partial_\lambda \phi^c\partial^\lambda\phi^d}} \,, }
where $\iota\sqrt{\cA_a\cA_b\dot\phi^a\dot\phi^b} = \cA_a\dot\phi^a$. For homogeneous backgrounds the energy density and pressure are
\begin{align}\label{psolf}
\varrho = \cV(\phi), \quad p = -\cV(\phi) - \cA_a\dot\phi^a,
\end{align}
as found in eqs.\ \eqref{eqn:energy} and \eqref{pressure}.  The scalar field equations following from eq.\ (\ref{act1}) similarly are
\begin{align}
\frac{1}{\sqrt{-g}}\partial_\mu\left( \frac{\iota\sqrt{-g}\;\cA_a\cA_b\partial^\mu\phi^b}{\sqrt{-\cA_c\cA_d\partial_\nu \phi^c\partial^\nu\phi^d}} \right) + \cV_{,a} - \frac{\iota\cA_{b,a}\cA_c\partial_\mu\phi^b\partial^\mu\phi^c}{\sqrt{-\cA_d\cA_e\partial_\nu \phi^d\partial^\nu\phi^e}} = 0 \,.
\end{align}
Once specialized to spatially homogeneous solutions these become
\begin{align}\label{fredeom}
-\frac{1}{\sqrt{-g}}\partial_0\left(\sqrt{-g}\cA_a \right) + \cV_{,a} + \cA_{b,a}\dot\phi^b = 0, \quad
\text{ or }
-3H\cA_a +\cF_{ab}\dot\phi^b + \cV_{,a} = 0 \,,
\end{align}
identical to eq.\ \eqref{Seom1}.

Because the target space metric $\widetilde{\cG}_{ab} = \cA_a\cA_b$ is constructed from a  product of a vector with itself, it has rank one. Consequently there is only one scalar fluctuation that appears in the kinetic term regardless of the nominal dimension of the target space, leaving all but one of the $\phi^a$ as auxiliary fields. Therefore, despite involving multiple scalars in its formulation, magnon inflation is effectively a single-clock theory and so gives only adiabatic perturbations (as we demonstrate explicitly further on). We remark in passing that one can also recover the standard canonical and derivatively coupled (k-inflationary) class of inflationary models through appropriate choices for $\cV$ and $\cA_a$ after integrating out the auxiliary variables. 

\subsection{Background evolution for magnon inflation through two-field examples}

It is instructive to see how the background evolution responds to choices made for the target-space quantities $\cV$ and $\cA_a$, so we next explore in more detail the field evolution in a few illustrative two-field examples. Of particular interest are the circumstances under which the evolution allows (or forbids) transitions between different signs for $\dot H$ (and so also for $\varepsilon$). 

\subsubsection{Relation to Chromo-natural inflation}\label{sec:ch}

We start by showing how the Chromo-natural inflation model \cite{Adshead:2012kp} motivates the single-derivative terms considered here. 
In this model, the inflaton is a gauge potential for a gauge group that contains an $O(3)$ factor, with an inflationary vev, $\tilde\psi$, that preserves invariance under simultaneous rotations in physical and gauge space: $A^b_i \propto a \tilde \psi \, \delta^b_i$ where $a = a(t)$ is the scale factor $b = 1,2,3$ is a gauge index while $i = 1,2,3$ counts spatial coordinates. The preferred frame is the one within which this residual rotational invariance is defined. This symmetry breaking pattern differs from the one related to $U^\m$,  and
the fluctuation spectrum is different in both theories. Still, for the aligned background both possibilities are equivalent and share the same slow-roll analysis (in particular the possibility to relax the conditions on the steepness of the potential).  

To make this more concrete, recall that in addition to the gauge potential, $\tilde\psi$, Chromo-natural inflation also has an axion field, $\tilde\chi$. For slow motion, the background evolution of these fields in FRW spacetime is dominated  by the action \cite{Adshead:2012kp, Adshead:2012qe, Martinec:2012bv, Adshead:2013qp, Dimastrogiovanni:2012st,Adshead:2013nka}:
\begin{align} \label{CNLag}
S_{CN} =  \int \exd^4x \,a^3 \left\{ \left[ \frac3{2a^2} \left[ \frac{\partial(\tilde \psi a)}{\partial t}\right]^2  - \frac{3\tilde{g}^2}{2} \, \tilde\psi^4 \right] 
+ \frac{\dot{\tilde \chi}^2}2 - \tilde\mu^4\left[ 1 + \cos\left(\frac{\tilde\chi}{\tilde f}\right)\right] - \frac{3\tilde{g}\lambda}{\tilde f} \left( \frac{ \tilde\chi \, \tilde\psi^2}{a} \right) \frac{\partial(\tilde\psi a)}{\partial t}  \right\} \,,
\end{align}
where the terms involving only $\tilde\psi$ come from the Maxwell action while those with only $\chi$ descend from the axion kinetic term and potential. The $\tilde\psi-\chi$ interaction starts life as an axion-$F\wedge F$ term (with coupling $\lambda$). Here $\tilde f$ is the axion decay constant and $\tilde\mu$ is the scale of explicit axion symmetry breaking while $\tilde g$ is a gauge coupling of the underlying theory. This can be put into the form of the Lagrangian considered here by integrating the last term by parts and normalizing the fields appropriately, so that
\begin{align}\label{CNLagII}
S_{CN} \simeq  \int \exd^4x\, a^3 \left\{  - \frac{3\tilde{g}^2F^4}{2} \, \psi^4   - \tilde\mu^4\left[ 1 + \cos\left(\frac{F\chi}{\tilde f}\right)\right] 
+ \left( \frac{\tilde{g}\lambda}{\tilde f} \right)F^4 \psi^3\dot\chi+\frac{3F^2}{2a^2} \left[ \frac{\partial(\psi a)}{\partial t}\right]^2 +  \frac{F^2\dot{\chi}^2}2 \right\} \,.
\end{align}
From this we read off the zero- and one-derivative components: $\cB = 0$ while
\bea
\cA_a \,\exd\phi^a =  \cA(\psi) \, \exd\chi &=& - \left( \frac{\tilde{g}\lambda F^4}{\tilde f} \right) \psi^3 \, \exd\chi, \quad \quad 
\cF_{\chi\psi} = -\cF_{\psi\chi} = \frac{3\tilde{g}\lambda F^4}{\tilde f}\psi^2,  \nn \\
\textrm{and} \quad \cV(\psi,\chi) &=& \frac{3\tilde{g}^2F^4}{2} \, \psi^4  + \tilde\mu^4\left[ 1 + \cos\left(\frac{\chi}{\tilde f}\right)\right]  \,. \label{eq:potch}
\eea
The scales $\m$ and $m$ read from these potentials can be small provided the gauge coupling satisfies $\tilde g\ll 1$. In this model $\cA_a$ is not parallel to $\partial_a \cV$. As a result, 
\be
\varepsilon = - \frac{ \dot\phi^a}{2 H} \, \partial_a \ln \cV = -\frac{3\tilde \cF^{ab}\cA_b\partial_a\cV}{2\cV} 
                   = \frac{3\tilde{g}^2 F^4\psi^4}{\cV} ,
\ee
up to second-derivative corrections, in agreement with what was found in \cite{Adshead:2012kp}. 

\subsubsection{Two-field example with nonzero $\partial_a \cV$ but vanishing $\varepsilon$}
\label{sec:twofieldI}

Consider the simplest two-field example with fields $\psi = \phi^1$ and $\chi = \phi^2$ with the choices
\be
  \cV = \cV(\psi) \quad \hbox{and} \quad \cA_a \exd \phi^a = \cA_1 \exd \phi^1 = \cA(\chi) \exd \psi  \,,
\ee
so $\cF_{12} = - \cF_{21}  =  -\cA'(\chi)$. This yields $\widetilde \cF^{12} = - \widetilde \cF^{\,21} = 1/ \cA'(\chi)$. The action becomes
\be
 S =  \int \exd^4x \sqrt{-g} \; \left\{\frac{\mpl^2}2 \, R - \cV(\psi) - \cA(\chi) \, U^\mu \partial_\mu \psi - \xi \, \Bigl( g_{\mu\nu} U^\mu U^\nu - 1 \Bigr)  \right\} \,.
\ee

The equations of motion obtained by varying $\xi$ and $U^\m$ are eqs.\ \pref{zetaeom} and \eqref{Ueom}. 
The aligned solution yields 
\be \label{Seom1x}
 \dot \chi = \frac{   \cV'(\psi) -3H  \cA(\chi) }{\cA'}  \qquad \hbox{and} \qquad
  \dot \psi  = 0  \,.
\ee
These field equations can be integrated explicitly to give the solutions (assuming $\cA' \ne 0$)
\be\label{eq:twofieldIsol}
 \psi = \hat\psi \qquad \hbox{and} \qquad 
 \cA(\chi(t)) = \frac{\hat\cV'}{3 \hat H} + \left[ \cA(\chi_0) - \frac{\hat \cV'}{3 \hat H} \right] \, e^{-3 \hat H(t - t_0)} \,,
\ee
where `hats' indicate evaluation at the constant value $\psi = \hat\psi$ (e.g.\ $\hat \cV' = \cV'(\hat \psi)$) while subscript `0' indicates evaluation at the initial time ({ e.g.} $\chi_0 = \chi(t = t_0)$).  This solution shows how $\chi$ relaxes to a steady-state value, $\chi_\infty$, with $\cA(\chi_\infty) = \hat\cV' /3\hat H$, on timescales of order the Hubble time. Because the equations are first order a constant force, $\cV'(\hat\psi)$, determines the late-time value of $\chi$ rather than the late-time field velocities.

Notice that because $\cA_a$ and $\partial_a \cV$ are chosen parallel in this example it follows that 
\be \label{vanish}
\widetilde \cF^{ab}  \cA_a \partial_b V = 0 ,
\ee
vanishes identically. As a consequence $\varepsilon$ and $\xi$ also vanish (cf. eqs.\  \eqref{zeta0eq} and \eqref{epsilon1stderiv}). 
This achieves the vacuum equation of state --- and so also a de Sitter gravitational geometry --- for free; a result that is trivially consistent with the equation of motion for $\psi$ whenever $\cA' \ne 0$ because eq.\ \pref{Seom1x} implies $\dot \psi = 0$ and so ensures that $\cV(\psi)$ remains constant along the flow lines of $U^\mu$. Notice that $\varepsilon = 0$ is true for all $t$ along these trajectories, even as $\chi$ rolls towards $\chi_\infty$. Most importantly, this is true {\em regardless} of the size of the slope, $\cV'(\psi)$, at the field-point of interest. We note in passing that $\xi=0$ means that $U^\m$ is not specified from the equations of motion. The aligned configuration is then a choice of `initial condition' which is compatible with the higher order corrections (cf.\ section \ref{sec:twoder}).

\subsubsection{Potentials  with linear $\cA_a$}\label{sec:lin}

Let us now consider the case of gaussian kinetic terms, for which $\cA_a$ is linear so that
\begin{align}
   \cA_a \exd\phi^a =  -\lambda_\phi \mu^3 \chi \, \exd\phi + \lambda_\chi \mu^3  \phi \, \exd\chi \,,
\end{align}
where $\lambda_\phi$, $\lambda_\chi$, and the fields $\phi$ and $\chi$ are dimensionless. The target-space 2-form has components
\begin{align}
   -\F_{\chi \phi} = \F_{\phi\chi}  = (\lambda_\phi + \lambda_\chi)\mu^3  \,.
\end{align}
The case $\lambda_\chi = - \lambda_\phi$ gives $\cA = -\lambda_\phi \mu^3 \, \exd(\phi \, \chi)$ while $\lambda_\chi = \lambda_\phi$ gives the case where $\cA_a$ is `pure curl'. 

With these choices, one has generically $\xi\neq 0$ (and thus alignment) and  the background equations of motion  become
\bea
(\lambda_\phi +\lambda_\chi) \dot{\phi} &=& -3 H \lambda_\chi \phi +\frac{1}{\mu^3 } \frac{\partial \cV}{\partial \chi}, \nn\\
(\lambda_\phi +\lambda_\chi) \dot{\chi} &=& -3H \lambda_\phi \chi - \frac{1}{\mu^3 }\frac{\partial \cV}{\partial \phi} , 
\label{eq:dpdc}
\eea
and so the sign of $\varepsilon$ is driven by
\be
\mpl^2 \dot{H} = -\frac{1}{2}\A_a \dot{\phi}^a = \frac{\mu^3}{2} \left(  \lambda_\chi \, \phi \, \dot\chi - \lambda_\phi \, \chi \, \dot\phi \right)
= -\frac{1}{2(\lambda_\phi + \lambda_\chi)} \left( \lambda_\phi \chi \, \frac{\partial \V}{\partial \chi}+   \lambda_\chi \phi \, \frac{\partial \V}{\partial \phi} \right) \,.
\ee
Notice that this is not changed by a simultaneous change of sign for both the $\lambda\,$s. It is clear that this can take either sign, even if $\cV$ itself is strictly non-negative.

\subsubsection*{Evolution with $\varepsilon$ changing sign}

To explore further when $\varepsilon$ can change sign we specialize to rotation-invariant potentials that are functions only of the single variable $x = \phi^2 + \chi^2$. Then 
\be
\label{eq:dxdt}
\dot x=-6H(\lambda_\chi \phi^2+\lambda_\phi \chi^2)/(\lambda_\chi+\lambda_\phi),
\ee
and
\begin{align}
\varepsilon =  3 \left( \frac{\lambda_\chi \phi^2 +\lambda_\phi \chi^2}{\lambda_\phi+\lambda_\chi}\right) \frac{\V'}{\V} \,,
\end{align}
where $\cV' = \exd \cV/\exd x$. Potentials with negative slopes ($\V' < 0$) have $\varepsilon < 0$ and so violate the null energy condition while those with positive slopes ($\V' > 0$) do not. 

Suppose we have a potential with both signs of $\cV'$. Can the field equations evolve from one region to the other? It is fairly straightforward  to see that such crossing is {\em not} possible so long as $H= 0$ whenever $\V' = 0$. In this case, this point is a fixed point of the
cosmological evolution (see eq.\  \eqref{eq:dxdt} or  eq.\ \eqref{eq:dpdc}). However, if $\cV'$ passes through zero somewhere where $H \ne 0$ (due to $\cV \ne 0$, or possibly due to contribution to $H$ from other forms of matter) then the evolution can pass through to change the sign of $\varepsilon$. In this case, provided ${H} > 0$, the scalar field dynamics in the regime $\varepsilon < 0$ is such that the fields move towards the origin (where $x=0$, where $\varepsilon$ vanishes). This happens independent of the form of the potential. In the case of symmetry breaking potentials, e.g. 
\begin{align}
\V(x) =g  \(\phi^2 + \chi^2 - v^2\)^2+\V_0 \, ,
\end{align}
 the fields \emph{climb} up the potential towards its maximum. This leads to a static late-time de Sitter configuration. 
 
A more general linearized analysis of homogeneous configurations near $\varepsilon = 0$ backgrounds using the single-derivative action is provided in subsequent sections.

\subsubsection*{Purely gaussian systems}

The simplest possibility is provided by a quadratic potential 
\be
   \cV = \frac12 \left( m_\phi^4 \phi^2 + m_\chi^4 \chi^2 \right) \,,
\ee
the scalar field equations are linear 
\begin{align}
\label{twofield_eom}
\[\begin{matrix} \dot{\phi} \\ \dot{\chi} \end{matrix}\] = \frac{1}{\lambda_\phi + \lambda_\chi}\[\begin{matrix} - 3H \lambda_\chi &  m^4_{\chi}/\mu^3 \\ -m^4_{\phi}/\mu^3 & -3 H \lambda_\phi \end{matrix}\]\[\begin{matrix} {\phi} \\ {\chi} \end{matrix}\]\, , 
\end{align}
and so can be solved explicitly.  Notice that the case $\lambda_\phi + \lambda_\chi = 0$ is singular, because in this case the kinetic term reduces to a total derivative when $H=0$. For this reason we assume this sum does not vanish, and so that $\cF_{ab} \ne 0$ (giving dynamics requiring at least two scalar fields).

From eq.\ \eqref{eq:dxdt} we know that the fields will roll to the point $\phi = \chi =0$.
To understand the nature of the solutions, let us point out that for the quadratic potential
\be \label{eqn:epsH}
\varepsilon = \frac{3\A_a \dot{\phi}^a}{2\V} = \frac{3}{\lambda_\chi+\lambda_\phi} \left( \frac{m_{\phi}^4 \lambda_\chi \phi^2 +m_\chi^4\lambda_\phi \chi^2}{m_{\phi}^4\phi^2 +m_\chi^4\chi^2}\right) \,,
\ee
which is  always positive (provided both $\lambda_a$'s share the same sign and both $m^2_a$'s share the same sign). The result reduces to $\varepsilon = \frac32$ when $\lambda_\phi = \lambda_\chi$, and this case is similar to matter domination.

\begin{figure}[h]
\begin{center}
\includegraphics[width = \textwidth]{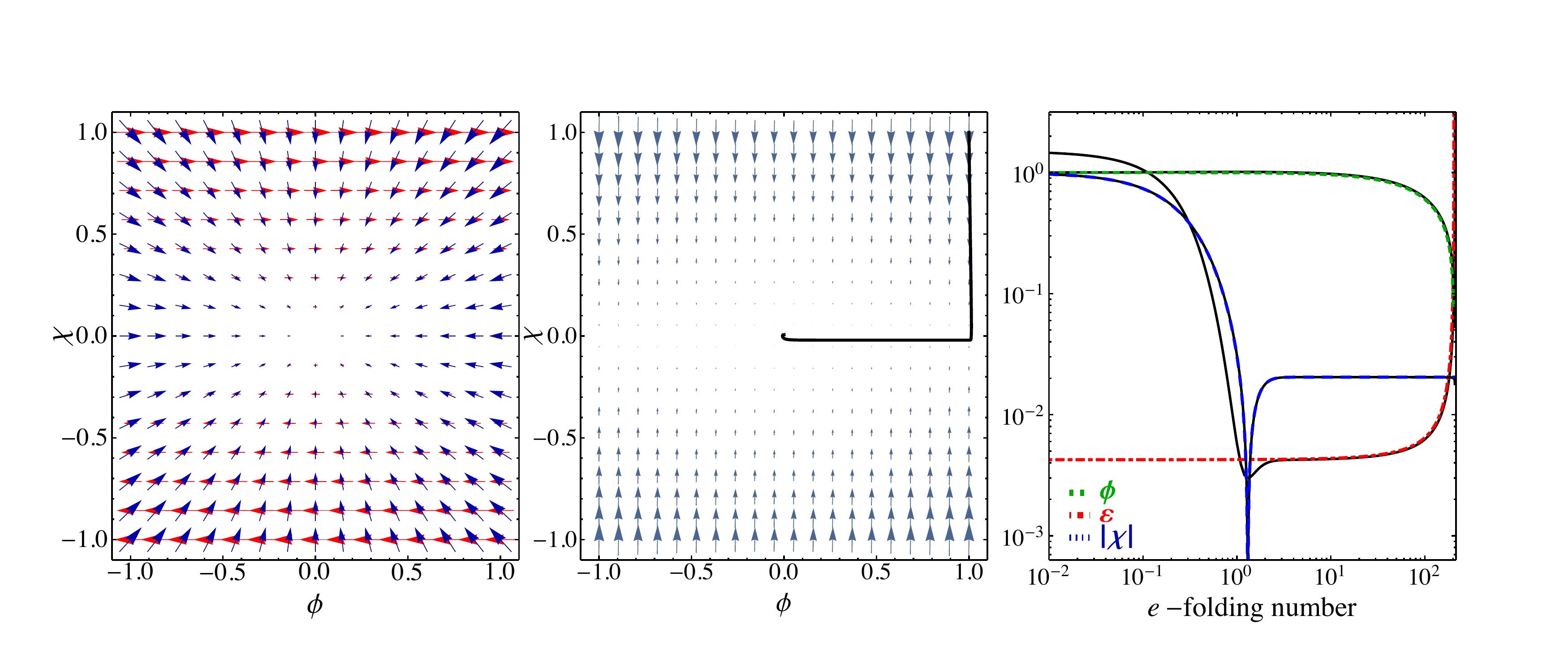}
\caption{Left panel: blue radially directed arrows denote the direction and size of $\V_{,{a}}$ while the red horizontal arrows give the direction and size of $\A_a$. Middle panel: arrows show $(\dot{\phi}, \dot{\chi})$ as predicted by the field equations, with the black line giving the trajectory followed by a homogeneous field evolving towards the origin from a specific initial condition. Most of the time is spent on the horizontal section of black curve, where the spacetime geometry inflates. Right panel: The evolution of the fields $\phi$ and $\chi$ is shown, as well as the evolution of the slow-roll parameter $\varepsilon$. In all cases, the numerical results are shown in black curves, while dashed green shows the analytic result of eq.\ \eqref{eqn:fieldsolsphi} for $\phi$, dotted blue shows eq.\ \eqref{eqn:fieldsolschi} for $\chi$ and dot-dashed red shows  eq.\ \eqref{eqn:epsilonapprox} for $\varepsilon$. In all panels, parameters are chosen to be $\lambda_{\phi} = 10$, $\lambda_{\chi} = 0.01$, $\mu = 0.01~ \mpl$, $m =0.0005 ~ \mpl$.}\label{fig:2field}
\end{center}
\end{figure}

Inflationary solutions are  possible when $\lambda_\chi \ne \lambda_\phi$. Let us  take for simplicity $m_{\phi} = m_{\chi} =: m$ and assume the hierarchy $\lambda_{\phi}\gg \lambda_{\chi}$. In this case evolution in the $(\phi,\chi)$ plane very generically passes through an inflationary solution on its way to the global minimum at $\phi = \chi = 0$, as may be seen in the left and middle panels of  figure \ref{fig:2field}.
Starting from its initial condition the system rolls quickly towards a region where $\cA_a$ and $\cV_{\,,a}$ are close to parallel ($\phi$ sits at its initial position while $\chi$ rolls down to $\chi^2 \ll \phi^2$).
During this time the slow-roll parameter $\varepsilon$ becomes small, indicating that the spacetime geometry is inflating. Eventually $\phi$ rolls off towards its minimum after which the system performs damped oscillations about the potential minimum  $\phi = \chi = 0$.
 
To understand the inflationary stage of the solution analytically, consider the regime  $\phi^2/\chi^2 \gg 1$, $\lambda_\phi/\lambda_\chi \gg 1$, in which case eq.\ \pref{eqn:epsH} becomes $\varepsilon \simeq 3 \lambda_\chi/\lambda_\phi$, while the Friedmann equation reads $6 \mpl^2H^2 \simeq m^4\phi^2$.  After switching to $e$-folding number,  the equations of motion  eq.\ \eqref{twofield_eom} read
\begin{align}
\frac{\exd{\phi}}{\exd N} = & -3 \frac{\lambda_\chi}{(\lambda_\phi +\lambda_\chi)} \phi + \frac{ \sqrt{6}  }{(\lambda_\phi +\lambda_\chi)}\frac{\mpl m^2}{\mu^3}\frac{\chi}{\phi},\\
\frac{\exd{\chi}}{\exd N} =  & -3\frac{ \lambda_\phi}{(\lambda_\phi +\lambda_\chi)} \chi - \frac{\sqrt{6}}{(\lambda_\phi +\lambda_\chi)}\frac{\mpl m^2}{\mu^3}.
\end{align}
These equations have solution, which can be found by direct integration
\begin{align}\label{eqn:fieldsolsphi}
\phi  = & \phi_0 \sqrt{e^{ -6 \frac{ \lambda_\chi N}{(\lambda_\phi +\lambda_\chi)}}\(1+\frac{2M_p^2m^4}{3 \mu^6 \lambda_\phi \lambda_\chi \phi_0^2}\)-\frac{2\mpl^2m^4}{3\mu^6 \lambda_\phi \lambda_\chi \phi_0^2}},\\\label{eqn:fieldsolschi}
\chi 
=  & \( \chi_0 + \frac{\sqrt{6}\mpl m^2}{3\mu^3 \lambda_\phi}\)e^{-3\frac{\lambda_\phi N }{(\lambda_\phi +\lambda_\chi)}} -  \frac{\sqrt{6}\mpl m^2}{3\lambda_\phi \mu^3},
\end{align}
where $\chi_0$ and $\phi_0$ are the initial values of the field, which we take to be $\phi_0 = 1$ and $\chi_0 = 1$ in the numerical example shown in figure \ref{fig:2field}.
Note that the evolution of $\chi$ is governed by $\lambda_\phi/(\lambda_\phi+\lambda_\chi) \approx 1$. Thus the solution for $\chi$ quickly becomes a constant, with $\dot{\chi} \approx 0$. The evolution of $\phi$ is governed by $\lambda_\chi/(\lambda_\phi+\lambda_\chi) \ll 1$, and is slow.

A more precise value for  $\varepsilon$ evaluated on this solution is
\be\label{eqn:epsilonapprox}
\varepsilon \simeq \frac{3}{\lambda_\phi}\(\lambda_\chi + \frac{2m^4\mpl^2}{3\mu^6 \lambda_\phi}\frac{1}{\phi^2}\) \,,
\ee
and thus, we see that in order to get an inflationary solution, for order unity fields, we require
\begin{align}
\frac{m^2\mpl}{\mu^3} \ll \lambda_\phi.
\end{align}
When this condition is satisfied, initially $\varepsilon$ is suppressed by $\lambda_\chi/\lambda_\phi$ but then begins to evolve as $\phi$ shrinks towards the origin. Using the solutions for the fields, we can find where inflation ends and estimate the number of $e$-folds
\begin{align}
N
 \approx & \frac{\lambda_\phi+\lambda_\chi}{6\lambda_{\chi}}\ln\[\frac{\(\phi_0^2+\frac{2\mpl^2m^4}{3 \mu^6 \lambda_\phi \lambda_\chi }\)}{2\frac{m^4 \mpl^2}{\mu^6 \lambda_\phi^2}\(1 + \frac{\lambda_\phi}{3\lambda_\chi}\)}\].
 \end{align}
Taking the log to be order $1$, we see that a hierarchy or the order of $\lambda_{\phi}/\lambda_{\chi}\sim\mathcal{O}(10^2)$ is required to get 60 $e$-foldings of inflation. 

There are several points worth emphasizing about the naturalness of such a hierarchy of parameters. First, although radiative corrections can change the size of the ratio $\lambda_\phi/\lambda_\chi$, the required tunings are at the level of 1 part in 100. Such hierarchies arise within the context of Chromo-natural inflation, and more generally hierarchies at the percent level are not particularly bothersome. (For a recent discussion of this point see for example \cite{Wells:2013tta}.) The real progress relative to standard two-derivative inflationary models is the absence of a condition demanding terms in the action (such as the inflaton mass) be smaller than Planck-suppressed quantities like the Hubble scale $H$. In this regard single-derivative terms are qualitatively different from the scalar potential because (unlike the scalar potential) they can be excluded by symmetries such as time-reversal invariance or unbroken Lorentz-invariance, and this allows their overall scale to be naturally hierarchically different from others UV scales in the problem (like $M_p$).

\section{Fluctuations}
\label{sec:stability}

The previous section shows that it is sometimes possible that $\varrho + p$ is negative, and so nominally violates the NEC. Although this need not in general imply an instability (see, for example, \cite{Rubakov:2014jja}), it behooves one to examine whether it does in any particular instance. In this section we present two calculations that suggest that instabilities need not be present. This is easily shown for $\epsilon >0$ while the analysis of the cases $\epsilon \leq 0$ is more involved. 

In the first calculation we examine homogeneous perturbations around the static backgrounds. We pay special attention to the cases for which $\varepsilon$ vanishes identically, since these are the most counter-intuitive. In the second calculation we consider inhomogeneous fluctuations which are also of intrinsic interest for the purpose of connecting to observations. We restrict this part of the fluctuation analysis to backgrounds for which $\varepsilon \ne 0$, however, to avoid a degenerate limit for curvature perturbations.\footnote{Although homogeneous instabilities can in certain situations be good since they might describe the evolution of the background away from one kind of cosmology (perhaps inflation) towards a different later-time attractor (and thereby perhaps end inflation), instabilities for non- zero momenta are more problematic (particularly if the instability can be made worse simply by increasing $k$).}

\subsection{Homogeneous fluctuations about $\varepsilon = 0$ backgrounds}
\label{sec:subdomderiv}

This subsection explores homogeneous fluctuations about the backgrounds for which $\varepsilon = 0$ identically; first showing how the single-derivative interactions are only marginally stable and then tracking how two-derivate scalar interactions stabilize or destabilize the leading marginal result. The goal is to identify what the leading contribution is to $\varepsilon$ and $\eta$ in the marginally stable case where $\varepsilon$ vanishes identically at the one-derivative level.

\subsubsection{Linearization of the single-derivative action}

We start by considering the presence of homogeneous perturbations $\delta \phi^a$ over given solutions  
\be
\label{eq:perturb}
  \phi^a = \varphi^a + \delta \phi^a \,,
\ee
and the aligned solution $U^\mu = u^\mu$. 
We linearize the field equations \eqref{Seom} in $\delta \phi^a$, leading to the linearized scalar field equation
\be \label{lineq1}
 \cF_{ab} (\varphi) \, \delta \dot \phi^b + \cM_{ab}(\varphi) \, \delta \phi^b = 0 \,,
\ee
with
\be
\label{eq:Mab}
 \cM_{ab} := \cV_{,ab} + \cF_{ac,b} \,\dot \varphi^c - 3 H_{,b} \, \cA_a  - 3 H \cA_{a,b}  \,.
\ee

Our interest is when $\cF_{ab}$ has an inverse, $\widetilde \cF^{ab}$, in which case \pref{lineq1} has the form
\be \label{lineartildeeq0}
    \delta \dot \phi^a + {\widetilde {\cM}^a}{}_{c} \, \delta \phi^c = 0 \,,
\ee
with ${\widetilde{ \cM}^a}{}_c := \widetilde \cF^{ab} (\varphi) \cM_{bc}(\varphi)$ given (after use of the background field equation) by
\be \label{widehatM}
 {{ \widetilde\cM}^a}{}_b = \left. \left[ \widetilde \cF^{ac} \left( \cV_{,c} - 3 H \cA_c \right) \right]_{,b} \right|_{\phi = \varphi} \,.
\ee

This has general solutions
\be
 \delta \phi^a(t) = {\left[ \cT \exp \left( - \int_0^t \exd \tau \,  \widetilde\cM (\tau) \right) \right]^a}_b \; \phi^b(0) \,,
\ee
where $\cT$ denotes the time-ordering of the exponential. Given that the 
background solution in \eqref{eq:perturb} is assumed to roll slowly, we can identify the solution as stable when the eigenvalues of $\widetilde\cM^a{}_b$ are strictly positive, marginally stable when they are non-negative and unstable once a negative eigenvalue exists. For the marginally stable eigenvalues,  their ultimate fate depends on the two-derivative interactions (more about which below). Of course this type of linearized analysis cannot identify the endpoint in the case of instability, but it can be used to self-consistently identify the {\em absence} of slow-scale instabilities.\footnote{A one-derivative analysis cannot rule out faster instabilities, but because these necessarily occur with characteristic rates $\Gamma \gg H$ they can be sought using  stability analyses from two-derivative interactions and neglecting the cosmological expansion.}  

Eq.~\pref{widehatM} gives the order of magnitude of the relaxation (or instability) rate, $\Gamma_r$, dictated by the magnitude of the eigenvalues of $\widetilde\cM^a{}_b$. These generically involve competing contributions from terms of order $\widetilde \cF^{ab} \cV_{,b} \sim (m^4/\mu^3)(v'/\alpha') \sim (m^2 H \mpl/\mu^3)(v'/\alpha')$ and those of order $H \widetilde \cF^{ab} \cA_b \sim H(\alpha/\alpha')$. Assuming $\alpha(\varphi)$ and $v(\varphi)$ are order unity we see that there are two generic cases: 
\begin{itemize}
\item If $m^2 \ll \mu^3/\mpl$ then we expect  $\Gamma_r \sim H \sim m^2/\mpl \ll \mu^3/\mpl^2$.
Recall that these scales are well-described by the single-derivative analysis (cf. eq.\ \eqref{Ebounds}).

\item If $m^2 \gg \mu^3/\mpl$ then some eigenvalues can be of order $\Gamma_r \sim m^4/\mu^3 \sim (m^2 \mpl/\mu^3) H \gg H$. This can still be within a single-derivative regime provided eq.\ \pref{Ebounds} remains satisfied. A rate of order $\Gamma_r \sim m^4/\mu^3$ can also easily lie within the domain of single derivatives, as is again most easily seen in the case $m \ll \mu \sim F \sim M \lsim \mpl$.
\end{itemize}

In both cases, once we made sure that we can trust the analysis based on eigenvalues of eq.\ \eqref{eq:Mab}  we don't see any obstruction to construct potentials that generically have no unstable directions.  Let us now see that the presence of marginal eigenvalues is
a relatively generic feature of the case $\varepsilon =0$. To understand the time evolution of these modes we need to include higher derivative operators, which we do in section \ref{sec:twoder}. This is also important to test the validity of the aligned solution.
 
\subsubsection{Marginal eigenvalues for static solutions and $\varepsilon=0$ }

We focus now in the case of static background solutions $\dot \varphi^a = 0$, which  requires 
\be
\label{staticcond}
\cV_{,a}(\varphi) - 3H \cA_a(\varphi) = 0.
\ee 
Then, the trace of the matrix  $\widetilde\cM^a{}_b$ reads
\bea
  \widetilde\cM^a{}_a = \left. -3 \widetilde \cF^{ab}  \left(  H \cA_b \right)_{,a} \right|_{\phi = \varphi} = \left. \widetilde \cF^{ab}  \left( \frac{ \cA_a \cV_{,b}}{2H \mpl^2} + \frac{3H \cF_{ba}}{2} \right) \right|_{\phi = \varphi} 
 &=& H \left(  \varepsilon + \frac{3N}{2}  \right) \,,
\eea
where $N$ is the number of scalar fields participating in the first-derivative term. Notice that this gives $3 H$ in the case of two scalars with $\varepsilon = 0$, in agreement with the explicit solution for the two-field model eq.\ \eqref{eq:twofieldIsol} (see also \eqref{twofield_eom}).

We now show that the existence of a zero (left-) eigenvector for $ \widetilde\cM$ is general whenever $\varepsilon$ vanishes for the case of static backgrounds and is not an accident of our two-field example of section \ref{sec:twofieldI}. 
We have seen that $\varepsilon$ vanishes identically whenever $\cA_a$ and $\cV$ are chosen so that $\widetilde \cF^{ab} \cA_a \cV_{,b}=0$. The latter condition automatically  ensures
\be \label{alphabeta}
 \widetilde \cF^{ab} (\alpha \, \cV_{,a} + \beta \,H \cA_{a} )( \cV_{,b} - 3 H \cA_b ) = 0 \,,
\ee
for all fields and for arbitrary constants $\alpha$ and $\beta$. Differentiating this equation with respect to $\phi^a$ and evaluating at a static background --- for which \pref{staticcond} holds --- then ensures that 
\be \label{QEDzeromode}
 0 = \left. (\beta + 3 \alpha) H \widetilde \cF^{ac} \cA_a ( \cV_{,c} - 3 H \cA_c )_{,b} \right|_{\phi = \varphi} = (\beta + 3 \alpha) H \cA_a(\varphi) { \widetilde\cM^a}{}_b \,.
\ee
Using \pref{QEDzeromode} in \pref{lineartildeeq0} then shows that $\delta \phi^a$, in an expansion about such a static point, satisfies (assuming $H \ne 0$)
\be 
  \cA_a (\varphi) \, \delta \dot \phi^a  = 0 \,,
\ee
and so that fluctuations in the direction $\cA_a(\varphi) \delta \phi^a \propto \cV_{,a} (\varphi)\, \delta \phi^a$ in field space are only marginally stable when only one-derivative actions are considered.

\subsubsection{Two-derivative terms}\label{sec:twoder}
We now compute the stability of this marginal direction once subdominant two-derivative terms are included into the action. The most general two-derivative terms that can be added to the action (up to integrations by parts and field redefinitions\footnote{Among the terms that may be eliminated are those proportional to $U_\mu \nabla_\nu U^\mu = 0$, which follows from the normalization conditions on  $U^\mu$.}) are
\bea \label{DeltaS}
 \Delta S &=&  \int \exd^4x \sqrt{-g} \; \Delta L \nn\\
 \hbox{with} \quad -\Delta L &=&  \frac12  \Bigl[ \cG_{ab}(\phi) \, g^{\mu\nu} + \cI_{ab}(\phi) \, U^\mu U^\nu \Bigr] \, \partial_\mu \phi^a \partial_\nu \phi^b + \frac12  \Bigl[ \cC^{(1)}(\phi) \, \nabla^\mu U^\nu \nabla_\mu U_\nu  \\
 && + \cC^{(2)}(\phi) (\nabla \cdot U)^2  + \cC^{(3)}(\phi) \, \nabla_\nu U^\mu \nabla_\mu U^\nu  + \cC^{(4)}(\phi) \, U^\lambda \nabla_\lambda U^\mu U^\nu \nabla_\nu U_\mu \Bigr] \nn \\
 &&  + \cC^{(5)}_a (\phi)\, U^\nu \left( \nabla_\nu U^\mu \right) \partial_\mu \phi^a + \cC^{(6)}_a(\phi)\, U^\mu \left(\nabla \cdot U \right) \partial_\mu \phi^a  \,, \nn
\eea
where $\cG_{ab}$ and $\cI_{ab}$ are symmetric in $a \leftrightarrow b$. To avoid unnecessary clutter in this section we absorb the scale $F^2$ into $\cG_{ab}$ (and the Lorentz-breaking scale $M^2$ into the other coefficients) rather than writing them explicitly. When restricted to constant $\phi^a$ coefficients $\cC^{(1)}$ through $\cC^{(4)}$ correspond to the basis of operators used in \cite{Jacobson:2004ts} for the Einstein-Aether theory. Some of these terms are absent if $U^\m$ is hypersurface orthogonal (see e.g.\  \cite{Blas:2010hb} for the case of constant coefficients).
If these terms survive until late times, one can use different observations to constrain their values \cite{Blas:2014aca}.  We will ignore 
this possibility in the following.

These terms change the equation of motion for $U^\mu$ adding to the left-hand side of eq.\ \pref{Ueom} the amount
\bea \label{DeltaUeomeq}
\Delta (\hbox{\ref{Ueom}})_\mu  &=&  \cI_{ab}  \, \partial_\mu \phi^a \dot \phi^b + \cC^{(4)}\, \dot U^\nu \nabla_\mu U_\nu  + \, \cC^{(5)}_a \nabla_\mu U^\nu \nabla_\nu \phi^a \\
&&\left.\qquad + \cC^{(6)}_a (\nabla \cdot U ) \nabla_\mu \phi^a -\, \nabla_\nu \right. \Bigl\{ \cC^{(1)}\, \nabla^\nu U_\mu + \cC^{(2)}\delta^\nu_\mu \nabla \cdot U +  \cC^{(3)} \nabla_{\mu}U^{\nu} \nn \\
&& \qquad\qquad +\cC^{(4)} U^\nu \dot U_\mu +\cC^{(5)}_a\, U^\nu \nabla_\mu \phi^a  + \cC^{(6)}_a \delta^\nu_\mu U^\lambda \nabla_\lambda \phi^a \Bigr\} \,, \nn
\eea
where an overdot denotes an application of $U^\mu \nabla_\mu$.  The left-hand side of the scalar field equation, \pref{Seom}, similarly acquires the new terms
\be \label{DeltaSeomeq}
\Delta (\hbox{\ref{Seom}})  = \frac{\partial (\Delta L)}{\partial \phi^a} + \nabla_\mu \left\{\cG_{ab}\nabla^\mu \phi^b + \cI_{ab} \,U^\mu \dot{\phi}^b + \cC^{(5)}_a \, \dot{U}^\mu + \cC^{(6)}_a \, U^\mu \nabla \cdot U \right\} \,, 
\ee
where $\Delta L$ is as defined in \pref{DeltaS} and $\partial \Delta L/\partial \phi^a$ is meant to convey ordinary differentiation of the coefficient functions, $\cG_{ab}(\phi)$, $\cI_{ab}(\phi)$ and $\cC^{(I)}(\phi)$. 

The new terms in the stress-energy are  (see appendix \ref{app:DeltaTcalc} for details)
\bea \label{DeltaTeq}
 \Delta T^{\mu\nu} &:=&  \frac{2}{\sqrt{-g}} \, \left( \frac{\delta \Delta S}{\delta g_{\mu\nu}} \right)  \\
 &=& \Delta L \, g^{\mu\nu}  + \cG_{ab} \, \partial^\mu \phi^a \, \partial^\nu \phi^b + \cC^{(1)}  \left( \nabla^\mu U^\lambda \nabla^\nu U_\lambda - \nabla_\lambda U^\mu \nabla^\lambda U^\nu \right) - \cC^{(4)} \dot U^\mu \dot U^\nu \nn\\
 && \qquad\qquad + \frac12 \nabla_\lambda \left[ \left( J^{\lambda \nu} - J^{\nu\lambda} \right) U^{\mu} + \left( J^{\lambda \mu} - J^{\mu\lambda} \right) U^{\nu} + \Bigl( J^{\mu\nu} +  J^{\nu\mu} \Bigr) U^\lambda \right] \,, \nn
\eea
where 
\be
 {J^\mu}_\alpha = {K^{\mu\nu}}_{\alpha \beta} \nabla_\nu U^\beta + \cC^{(5)}_a \, U^\mu\partial_\alpha \phi^a + \cC^{(6)}_a \, \dot \phi^a \, \delta^\mu_\alpha \,,
\ee
with
\be
 {K^{\mu\nu}}_{\alpha\beta} := \cC^{(1)} \, g_{\alpha \beta} \, g^{\mu\nu} + \cC^{(2)} \, \delta^\mu_\alpha \, \delta^\nu_\beta + \cC^{(3)} \, \delta^\mu_\beta \, \delta^\nu_\alpha + \cC^{(4)} \, g_{\alpha \beta} U^\mu U^\nu \,.
\ee
One also needs to remember that the value of $\xi$ is corrected by the two-derivative terms.

\subsubsection*{Perturbative evaluation}

With the goal of evaluating perturbatively close to the solutions $\{ u^\mu, \varphi^a, \hat H^2 = \hat\cV/3\mpl^2 \}$ of the previous section, we next evaluate eqs.~\pref{DeltaS} through \pref{DeltaTeq} at these earlier, zeroth-order, solutions. These are then the sources from which we compute the small changes 
\be
\label{eq:perturb_all}
 U^\mu = u^\mu + \delta U^\mu \,, \quad
 \phi^a = \varphi^a + \delta \phi^a \quad \hbox{and} \quad
 H = {\hat H} + \delta H \,. 
\ee

We start by evaluating \pref{DeltaUeomeq} at the background level,
\bea \label{DeltaUeomeqbg}
\left. \phantom{\frac12} \Delta (\hbox{\ref{Ueom}})_\mu \right|_{\rm bg} &=&  - \left[ \cI_{ab}  \, \dot\varphi^a \dot \varphi^b + 3  \left( \cC^{(1)} + \cC^{(3)} \right) {\hat H}^2 + 3{\hat H} \left( \cC^{(6)}_a - \cC^{(5)} \right) \dot \varphi^a \right] u_\mu \nn\\
&& \qquad  - \nabla_\mu \left[ \left( \cC^{(1)} + 3 \cC^{(2)} + \cC^{(3)} \right) {\hat H} + \cC^{(6)}_a \dot \varphi^a \right]   \nn\\
&& \qquad\qquad  - u_\mu \, u \cdot \nabla  \left[ \left( \cC^{(1)} + \cC^{(3)} \right) {\hat H} - \cC^{(5)}_a \dot \varphi^a \right]  \,.
\eea
This is compatible with  $\delta U^\mu \propto u^\mu$, and the normalization of $U^\m$ will follow from the change $\Delta\xi$ in the value of $\xi$. This 
finally shows that the aligned configuration is a solution. 

The change to the scalar field equation is determined by adding \pref{DeltaSeomeq} to \pref{Seom} and linearizing about the background solution, leading to the addition of an inhomogeneous term to \pref{lineq1}
\be \label{lineq}
 \cF_{ab} (\varphi) \, \delta \dot \phi^b + \cM_{ab}(\varphi) \, \delta \phi^b = \cJ_a(\varphi) \,,
\ee
with source term
\bea \label{Jdef}
 \cJ_a &:=&  \nabla_\mu \left[ \cG_{ab}\nabla^\mu \varphi^b + \left( \cI_{ab} \, \dot{\varphi}^b + 3{\hat H}\,\cC^{(6)}_a \right) u^\mu \right] + \frac{\partial (\Delta L)}{\partial \varphi^a}\nn\\
&=& - \cQ_{ab} \left( \ddot\varphi^b + 3 {\hat H} \, \dot \varphi^b + \Gamma^b_{cd} \, \dot \varphi^c \dot \varphi^d \right) + 3{\hat H} (\cC^{(6)}_{a,b} - \cC^{(6)}_{b,a} ) \dot \varphi^b \nn\\
&& \qquad\qquad - \frac{3{\hat H}^2}{2} \left[ ( \cC^{(1)} +3 \cC^{(2)} +  \cC^{(3)} )_{,a}  - 6 \,\cC^{(6)}_a \right]  \,,\nn
\eea
where $\Gamma^b_{cd}$ are the Christoffel symbol built from the target-space metric, $\cQ_{ab} := \cG_{ab} - \cI_{ab}$.

When $\cF_{ab}$ has an inverse,  \pref{lineq} becomes the inhomogeneous version of the linearized equation studied earlier
\be \label{lineartildeeq}
    \hat\cO \, \delta \phi^a := \delta \dot \phi^a + {{ \widetilde\cM}^a}{}_{c} \, \delta \phi^c = \widetilde \cJ^a \,,
\ee
where $\widetilde \cJ^a := \widetilde \cF^{ab} \cJ_b$. The general solution is a sum of a solution to the homogeneous part discussed in the previous section, plus any particular integral that includes the nonzero $\widetilde \cJ^a$. 

Our interest is when $\cA_a$ and $\cV$ are chosen so that the eigenvalues of ${{\widetilde \cM}^a}{}_b$ are strictly non-negative, so that the solutions to the leading-derivative equations are not unstable. The case of positive eigenvalues was discussed in  the previous sections, while the
evolution of marginal eigenstates requires the analysis of terms with higher order derivatives.  For this analysis we will focus in the
case of constant background fields. 
In this case any late-time rolling of the fields must be driven by the source term $\cJ_a$, so we first ask whether this (and $\Delta T^{\mu\nu}$) can be nonzero for static backgrounds satisfying $\dot \varphi^a = \dot {\hat H} = 0$ and $U^\mu = u^\mu$. Evaluating \pref{Jdef} with $\dot \varphi^a = \ddot \varphi^a = 0$ gives
\be
  \cJ_a =  - \frac{3{\hat H}^2}{2} \left[ ( \cC^{(1)} +3 \cC^{(2)} +  \cC^{(3)} )_{,a}  - 6 \,\cC^{(6)}_a \right]  \qquad \hbox{(when $\dot\varphi^a = \ddot \varphi^a = 0$)} \,,
\ee
which need not vanish. The directions $\eta_a $, where $\eta_a \widetilde \cM^a{}_b \ne 0$,
 decay to the late-time static solution: $ \delta \phi^a_\infty \sim (\widetilde \cM)^{-1} \widetilde \cJ$. 
For slow roll it is the zero eigenvectors, $\perp_a \widetilde \cM^a{}_b = 0$, that are of more interest, and these directions do exist --- with $\perp_a \propto \cA_a(\varphi) \propto \cV_{,a}(\varphi)$ --- for static solutions provided \pref{vanish} also holds (cf. \eqref{QEDzeromode}). For these directions the linearized field equations state
\be \label{one}
    \perp_a \delta \dot \phi^a = \perp_a  \widetilde \cJ^a = \widetilde \cF^{ab} \perp_a  \cJ_b \,,
\ee
and so for these the late-time solutions asymptote to constant {\em velocity} rather than constant position.

\subsubsection*{Slow-roll parameters}

We are now in a position to evaluate the perturbed slow-roll conditions for the case where the
leading contribution cancels. At linear order in the perturbations in \eqref{eq:perturb_all} these receive contributions from two sources: $(i)$ the linearized perturbations of the first-derivative slow-roll conditions, and $(ii)$ the contributions of $\Delta T^{\mu\nu}$ to the slow-roll conditions, evaluated at the zeroth order static background solution.

We start by evaluating $\Delta T^{\mu\nu}$, in which we must also take care to include the change to $\xi$ induced by the addition of $\Delta S$ to the action. Inspection of \pref{DeltaUeomeqbg} shows that the presence of $\Delta S$ for a static background shifts $\xi \to \xi + \Delta \xi$ with
\bea
 2 \Delta \xi &=& u^\mu  \Delta (\hbox{\ref{Ueom}})_\mu 
=  3  \left( \cC^{(1)} + \cC^{(3)} \right) {\hat H}^2  \,. \nn
\eea
Using this and \pref{DeltaTeq} the correction to the stress energy is then given by
\bea
 \Delta T^{\mu\nu} - 2 \Delta \xi \, U^\mu U^\nu 
 &=& \frac{3 {\hat H}^2}{2} ( \cC^{(1)}  + 3 \cC^{(2)} +  \cC^{(3)} ) g^{\mu\nu} \,.
\eea 
This is the standard result relating the terms $\cal C^{(I)}$ in \eqref{DeltaTeq} to a modification of the value of Newton's constant in Friedmann equation by ${\cal O}(M^2/M_p^2)$ \cite{Carroll:2004ai}.  This contribution does not affect the value of $\varepsilon$.

The leading corrections therefore come from the change in the leading order expressions due to the modification of the motion of the fields,
\be
 \varepsilon \simeq - \frac{ \delta \dot\phi^a}{2 {\hat H}} \, \left( \frac{\cV_{,a}}{\cV} \right)_{\phi = \varphi} \,,
\ee
with $\delta \dot \phi^a$ evaluated using \pref{one} (the stable fields asymptote to a constant). 

The order of magnitude of this result may be estimated by restoring the explicit dimensions of the coefficients, with $\cG_{ab} \sim F^2$, $\cA_a \sim \mu^3$ and $\cC^{(I)} \sim M^2$. This implies $\cJ_a \sim {\hat H}^2 M^2$ and so $\delta \dot \phi^a \sim \widetilde \cJ^a = \widetilde \cF^{ab} \cJ_b \sim {\hat H}^2 M^2/\mu^3$. The leading dependence of $\varepsilon$ on these scales therefore is
\be \label{vepsest}
 \varepsilon \sim \frac{{\hat H} M^2}{\mu^3}\frac{v'}{v} \sim \frac{m^2 M^2}{\mpl \mu^3}\frac{v'}{v} \,.
\ee
Applying the same estimates to $\eta = \dot \varepsilon/{\hat H} \varepsilon$ (and assuming, as above, all dimensionless functions of the fields and their derivatives are order unity) then predicts $\eta \sim \dot \phi/{\hat H} \sim \varepsilon\, v/v'$. When $\mu \sim M$ this implies (for
a potential with $v'\sim v$) $\eta \sim \varepsilon \sim {\hat H}/M \sim m^2/\mpl M$, which can be naturally small given a moderate hierarchy like $m \ll M \lsim \mpl$. 

We remark in passing that the assumption that the fields are order unity at the static solution implies a relation between the otherwise independent scales $m$ and $\mu$. This relation arises because static solutions require the background values to adjust so that $\cA_a$ and $\cV_{,a}$ satisfy $3 {\hat H} \cA_a = \cV_{,a}$, so if this is satisfied by order-unity field values then the statement that $\cA_a \sim \mu^3$ and $\cV \sim m^4$ implies the scales $m$ and $\mu$ must be related by 
\be \label{Hest}
 m^4 \sim {\hat H} \mu^3 \qquad \hbox{and so} \qquad m^2 \sim \frac{\mu^3}{\mpl}  \qquad \hbox{and} \qquad
 {\hat H} \sim \frac{\mu^3}{\mpl^2} \,.
\ee
This is  compatible with the bounds \eqref{Ebounds}.
Using this in the estimate \pref{vepsest} then implies $\eta \sim \varepsilon \sim (M/\mpl)^2$ is determined purely by the value of $M$, independent of $\mu$. This allows $M$ to be inferred directly from measurements of\footnote{The speed of gravitational waves is modified by the terms ${\cal C}^{(I)}$ by quantities of ${\cal O}(M^2/M_p^2)$, see e.g. \cite{Blas:2014aca}. For the
set-up we are currently considering, these effects are negligible.}  $r$ while \pref{Hest} relates the inferred value of $\mu$ to the Hubble scale during inflation. In particular, the observed amplitude of scalar fluctuations implies
\be
 \left( \frac{m^4}{\varepsilon} \right)^{1/4} \sim   \left( \frac{\mu^3}{M} \right)^{1/2}    \simeq 7 \times 10^{16} \; \hbox{GeV} \,.
\ee

For instance, if phenomenology were to tell us $\varepsilon \sim 10^{-2}$ (as would be implied by detection of primordial gravitational waves with $r$ close to present limits, for example), then this inflationary mechanism would indicate $M/\mpl  \sim \sqrt{\varepsilon} \; \sim 0.1$. Demanding the proper amplitude of scalar fluctuations then requires ${\hat H} / \mpl \sim 10^{-5} \sqrt{8 \pi^2 \varepsilon} \sim 10^{-5}$ and so $\mu \sim ({\hat H} \mpl^2)^{1/3} \sim 0.02 \, \mpl$. For scales $M$ much smaller than these values $r$ rapidly becomes undetectable.

\subsubsection*{Two-field example}

To see how this works more explicitly we return to the simple two-field example of section \ref{sec:twofieldI}. Now we also include generic two-derivative terms that depend on both fields, $\phi^a = \{ \psi , \chi \}$.  The lowest-order field equations for this system are given in \eqref{Seom1x},  
so static solutions require $3H \cA(\chi) = \cV'$.

Linearizing about a static background ($\psi = \hat \psi + \delta \psi$ and $\chi = \hat \chi + \delta \chi$ for which $\hat \cA'  \neq 0$), and including the two-derivative interactions gives the evolution equation for the would-be zero mode, $\delta \psi$:
\be
  \delta \dot \psi = \widetilde \cJ^\psi =  \frac{\hat \cJ_\chi}{\hat\cA'} =  \frac{3{\hat H}^2}{2\hat \cA'} \left[ ( \hat\cC^{(1)} +3 \hat\cC^{(2)} +  \hat\cC^{(3)} )_{,\chi}  - 6 \,\hat\cC^{(6)}_\chi \right]\,.
\ee
The leading two-derivative contribution to the slow-roll parameter $\varepsilon$ therefore becomes
\bea
 \varepsilon \simeq - \frac{\delta \dot \psi}{2 {\hat H}} \left( \frac{\hat\cV_{,\psi}}{\hat \cV} \right)  &\simeq& \frac{3{\hat H}}{4\cA'} \left[ ( \hat\cC^{(1)} +3 \hat\cC^{(2)} +  \hat\cC^{(3)} )_{,\chi}  - 6 \,\hat\cC^{(6)}_\chi \right] \left( \frac{\hat\cV_{,\psi}}{\hat \cV} \right) \nn\\
 &=&  \frac{9{\hat H}^2 \hat\cA}{4\hat\cV \cA'} \left[ ( \hat\cC^{(1)} +3 \hat\cC^{(2)} +  \hat\cC^{(3)} )_{,\chi}  - 6 \,\hat\cC^{(6)}_\chi \right]\,,
\eea
where the last equality uses the static relation between $\hat \chi$ and $\hat \psi$. 

This determines the sign and magnitude of $\varepsilon$ in terms of the sign and magnitude of $\cC^{(6)}$,  the $\chi$-derivatives of the coefficients $\cC^{(1)}$ through $\cC^{(3)}$ appearing in $\Delta L$, and the gradient of the scalar potential. Notice it is the marginally {\em unstable} solution that is desired if we wish eventually to exit inflation. 

\subsection{Cosmological fluctuations}
\label{sec:cosmicfluc}

In this section we discuss the action for quadratic scalar-metric fluctuations obtained by coupling the action at eq.\ \pref{act1} to gravity in comoving gauge.  For simplicity we  restrict ourselves to the case where there are only two fields. 

The mixing of scalar modes with the metric through gravitational interactions provides another way for scalar fluctuations to sample the two-derivative terms in the action, although this time it is those of the Einstein-Hilbert action rather than any explicit two-derivative scalar interactions present in the Lagrangian before coupling to gravity. These gravity-induced interactions are natural to examine, since they are self-contained and only depend on the original single-derivative scalar action (since metric perturbations are sourced primarily by the matter content that is driving the background evolution) together with the standard gravitational couplings already required to discuss inflation. 

We find that the curvature perturbations are generically adiabatic with a finite sound speed. Although their adiabatic, single-clock  character is most easily understood using the classically equivalent formulation of section \ref{sec:effdesc}, we verify that one reproduces the same results by directly perturbing the full theory of section \ref{sec:fulltheory}. As usual the fluctuations degenerate in the limit where $\varepsilon = 0$ due to the enhanced symmetry of de Sitter space. 

\subsubsection*{Parameterization of the fluctuations}

Following appendix \ref{app:MIPerts}, we write the Einstein-Hilbert action in ADM form \cite{Arnowitt:1962hi}
\begin{align}
S_{G} = & \frac{\mpl^2}{2}\int \exd^4 x N \sqrt{h}\[R^{(3)} + \frac{1}{N^2}(E^{ij}E_{ij} - E^2)\].
\end{align}
where the metric is given by
\begin{align}
\exd s^2 = -N^2 \exd t^2 + h_{ij}(\exd x^i + N^i \exd t)(\exd x^j + N^j \exd t).
\end{align}

For two scalar fields we are free to expand the scalar fields into background, $\varphi^a(t)$, plus fluctuations using
\begin{align}
\phi^a({\bf x}, t) = \varphi^a(t+\pi ({\bf x}, t) ) + \N^a(t+\pi ({\bf x}, t)) \;\sigma({\bf x}, t),
\end{align}
where $\pi({\bf x}, t)$ parameterizes a translation in time along the inflationary trajectory and $\sigma$ represents the isocurvature mode normal to the inflationary trajectory. Here the target-space vectors, $\T^a$ and $\N^a$ decompose fluctuations into directions tangent and normal to the inflationary trajectory according to
\be
\T^a := \frac{\dot\varphi^a}{\sqrt{\dot{\varphi}^c\dot\varphi_c}}, \quad \text{and} \quad
\N^a := \delta^{ac}\epsilon_{cb}\T^b \,.
\ee

We expand the lapse and shift as
\be
N = 1+\alpha_{1} +\ldots ,\quad \hbox{and} \quad
N^i =  h^{ij}\partial_j\theta_1+ \ldots,
\ee
where we keep only terms linear in the fluctuations, $\alpha = \alpha_1$ and $\theta = \theta_1$, since at quadratic order in the action second-order quantities like $\theta_2$ and $\alpha_2$ just multiply lower-order constraints \cite{Maldacena:2002vr}.

Finally, we drop the vector degrees of freedom altogether since we can show these to be zero\footnote{Vector degrees of freedom are present when the operators involving second derivatives of $U^\m$ \eqref{DeltaS} become relevant, as happens in Einstein-aether theory \cite{Jacobson:2008aj}.} when $\varepsilon \neq 0$ and $U^\m$ is fully aligned with the co-moving cosmic 4-velocity, $u^\mu$. The study of tensor modes on the other hand is unchanged by any of our new ingredients. 

\subsubsection*{Unitary (or comoving) gauge}

In unitary gauge, the inflaton is not perturbed along its trajectory and $\pi = 0$. We adopt coordinates to write the scalar component of the metric on the spatial hypersurfaces in the form
\begin{align}
h_{ij} \equiv a^2(t)e^{2\curv}\delta_{ij}.
\end{align}
In this gauge the fluctuations in the scalar field about the background, $\varphi^a(t)$, are strictly orthogonal to the inflationary trajectory
\begin{align}
\phi^a({\bf x}, t) = \varphi^a(t) + \N^a(t)\,\sigma(t, x) \,.
\end{align}
The physical scalar degrees of freedom are the curvature fluctuation, $\curv$, and isocurvature mode, $\sigma$.

With these definitions the unitary gauge quadratic action for fluctuations about the background solution is given by (for details see appendix \ref{app:MIPerts})
\begin{align}\nn
S^{(2)}  = &  \int \exd^3  x \, \exd t \; a^3 \Bigg[- \frac{\varepsilon \mpl^2 }{a^2}\delta^{ij}\partial_i\curv \partial_j \curv   + \frac{\A_\N}{H a^2 } \; \delta^{ij}\partial_{i}\partial_j\curv \,\sigma + \cF_{\N\T} \frac{\dot{\varphi} }{H} \; \sigma\,\dot\curv   
 \\
 &  - \(\M_{\N\N} -\M_{\T\T}+ \cF_{\N\T}\frac{\A_\N \dot{\varphi}}{2 H \mpl^2 }\)\frac{\sigma^2}{2}-  {\A}_\N \delta_1\tilde{U}^{i} \partial_i\sigma  +H^2 M_{\rm Pl}^2 \varepsilon \(a^2\delta_{ij}\delta_1 \tilde{U}^i \delta_1 \tilde{U}^j \)  
\Bigg]\label{eqn:unitact}
\end{align}
where $\delta_1\tilde{U}^i = N^i + \delta_1 U^i$ and $\cF := \cF_{\N\T} = \frac12 \epsilon^{ab} \cF_{ab}$ is the target-space field strength, while $\dot\varphi := \T^a \dot \varphi_a = \sqrt{\dot \varphi^a \dot \varphi_a}$ and $\cA_\N = \N^a \cA_a$, $\M_{\N\N} = \N^a \N^b \M_{ab}$ and $\M_{\T\T}= \T^a \T^b \M_{ab}$ with $\cM_{ab}$ as defined in \eqref{eq:Mab}
and $\cA_a$ and $\cM_{ab}$ evaluated at the background, $\varphi^a$. 

The first term of eq.\ \eqref{eqn:unitact} appears to hint at the presence of an instability when $\varepsilon < 0$, since the spatial gradient terms have the wrong sign. Furthermore, in Fourier space this instability (if present) appears to grow with momentum, $k$. However, in order to draw firm conclusions we need to explore the circumstances under which such an instability exists in a more careful manner.

\subsubsection*{Stability}

To clarify the stability issue we integrate out the quantities $\delta_1 \tilde{U}^i$ and $\sigma$, which appear in the above action purely as auxiliary fields. Notice that the isocurvature mode, $\sigma$, is undifferentiated here which can be understood from the degeneracy of the target-space metric in the classically equivalent action \pref{cusc}. When performing this integration we assume $\varepsilon \ne 0$, but do not assume it to be positive. We focus on the quadratic part of the action where the required functional integral is Gaussian, and is computed by evaluating it at the appropriate saddle point. This leads (in Fourier space) to the following quadratic action involving only the propagating fields, $\curv_k$,
\begin{align}\label{effact}
S^{(2)} = &\int \frac{\exd^3 k}{(2\pi)^3} \, \exd t \; a^3 \left\{ \frac{\dot{\varphi}^2\cF_{\N\T}^2}{2H^2 \Omega(k)} \; \dot{\curv}_k^2 - \mpl^2\[\frac{ L^4 \varepsilon}{\Omega(k)}  -\frac{1}{a}\frac{\exd}{\exd t}\(\frac{a\cA_\N}{2\mpl^2\Omega(k)}\frac{\dot\varphi \cF_{\N\T}}{H^2}\) \] \frac{k^2}{a^2}\; \cR_k^2  \right\}
\end{align}
Here
\begin{align}
\label{omegadef}
\Omega(k) := 
L^4 +\frac{\cA_\N^2}{2\varepsilon \mpl^2 }\frac{k^2}{a^2H^2} , \quad \quad 
\mathrm{with}\quad \quad 
L^4 :=  \(\M_{\N\N}-\M_{\T\T}+ \cF_{\N\T}\frac{\A_\N }{2 H \mpl^2 }\dot{\varphi}\) \,.
\end{align}
Because it involves only the propagating fields, eq.\ \eqref{effact} allows crisper statements about stability, including the following:
\begin{itemize}
\item The curvature perturbation is gapless, since all but the kinetic term vanishes when $k \to 0$. In particular, this means that on large scales $\curv$ is exactly conserved. Despite incorporating multiple fields, cosmological perturbations are adiabatic, showing that it behaves as a single-clock system. 
\item 
The kinetic term vanishes when $\cF_{\N\T} =0$ or $\dot{\varphi} = 0$, so curvature perturbations cease to propagate at finite speed in this limit.\footnote{Formally,  this corresponds to the limit where the sound speed of the curvature perturbations diverges.} When nonzero, the sign of the kinetic term is controlled by the sign of $\Omega(k)$, and so when $\Omega(k) < 0$ the theory contains ghosts. 
\item For small $k$ the leading gradient terms go like $k^2$ and we require them to be negative in $S^{(2)}$ and finite if we are to avoid instability towards the formation of spatial inhomogeneities. 
\end{itemize}
Stability requires the absence of both ghosts and gradient instabilities across all wavelengths of interest. Consider first ghosts: whether a ghost propagates or not is determined by the sign of the kinetic term in \pref{effact}. We first observe that when $L^4 < 0$, $\Omega(k) < 0$ as $k/(aH) \to 0$ implying an instability of the background at long wavelengths and so we focus in what follows on systems for which $\M_{\N\N} >\M_{TT}$ and these terms dominate the Planck-suppressed final term in the last line of \pref{omegadef}. In this case $L^4 > 0$ and so the kinetic term is always positive when $\varepsilon > 0$ (and so no ghosts in this case). 

The case where $\varepsilon < 0$ is less straight-forward. Modes with momenta smaller than
\begin{align}\label{eqn:ghost}
\frac{k^2}{a^2H^2} < \( \frac{2\abs{\varepsilon} \mpl^2}{\A_\N^2} \) L^4 \,
\end{align} 
are healthy whereas modes that violate the above inequality are ghost-like. In cases where $L\sim m$ (and if \eqref{Hest} is satisfied) this corresponds to wavelengths where our analysis neglecting two-derivative terms should apply. However, if $L$ is dominated by a large scale\footnote{An estimate from \eqref{effact} indicates an instability rate of order $\Gamma > \varepsilon H \left(M_p L^2/\mu^3\right)^{2}$.} this instability may evolve quickly enough not to be reliably described while neglecting two-derivative terms in the action. Similarly, if $\cA_\cN^2 < 2\abs{\varepsilon}L^4\mpl^2/F^2$ all modes are healthy that have momenta below the scale at which two derivative terms become relevant. Although one must check on a case-by-case basis, evidently the case $\varepsilon < 0$ need not imply ghost instabilities. 

Concerning gradient instabilities, at long wavelengths (i.e. when $\Omega(k) \to L^4$) we see from (\ref{effact}) that an instability sets in whenever
\begin{align}\label{tildeeps}
\tilde\varepsilon := \varepsilon   - \frac{1}{a}\frac{\exd}{\exd t}\(\frac{a\cA_\N\dot\varphi \cF_{\N\T}}{2\mpl^2H^2 L^4}\) < 0.
\end{align}
Note that it is possible for the above to be positive even when $\varepsilon < 0$.\footnote{In the case where $L\sim m$ and $\cA_\cN \neq 0$, it is possible for the second term on the right-hand side of eq.\ (\ref{tildeeps}) to dominate, implying the possible absence of long wavelength gradient instabilities even when $\varepsilon < 0$.} Away from the long wavelength limit one must consider the full expression within the square brackets of (\ref{effact}) and ensure that this contribution to the action is negative for all modes up the scale where two derivative terms  in the action (those neglected in \eqref{act1}) become relevant. Again instability must be checked on a case-by-case basis. A particularly simple special case of later interest is when $\cA_\cN = 0$, in which case avoidance of ghost and gradient instabilities requires $\varepsilon > 0$.

We finish this discussion with two remarks about \eqref{effact}. If we assume that these gravitationally induced two-derivative terms dominate all other two-derivative terms (a big if), then when $L\sim m$ and \eqref{Hest} holds the speed of propagation of $\cR$ is $c_s^2\sim \cO(\varepsilon)$ if the first term in the square brackets in \eqref{effact} dominates, while $c_s^2\sim \cO(1)$ if the second term dominates. When the perturbations are subluminal some obstructions to the existence of UV completion within Lorentz invariant set-ups may not arise \cite{Rubakov:2014jja}. 

For the case where the other two-derivative terms have coefficients, $\cC^{(I)}$, that do not depend on the scalar fields the conditions for stability have been studied elsewhere  \cite{Jacobson:2008aj,Blas:2014aca,Blas:2011en,Blas:2010hb}, as have aether-scalar couplings in the case of a single scalar field \cite{Donnelly:2010cr, Solomon:2013iza, Blas:2011en}. These studies show that stable solutions can exist, provided inequalities amongst the coefficients $\cC^{(I)}$ are satisfied. We leave the detailed study of how these cases generalize for arbitrary scalar couplings to a future study.

\subsubsection*{When $L^4$ dominates}

Things simplify somewhat if $L^4$ is larger that any other scale in the problem (except for  $\mpl$).  Such a situation arises, for instance, if $\cM_{\N\N} \gg F^4$. (Although this case is not completely generic since it requires an additional hierarchy in the potential sector, it is still an interesting possibility within the class of magnon inflation models.) In this case one can expand the above in a derivative expansion; keeping only the leading order then results in 
\begin{align}\label{csMI}
S^{(2)} = \int \frac{\exd^3 k}{(2\pi)^3} \, \exd t \; a^3 \[\mpl^2\tilde\varepsilon\left(\frac{\dot\calR^2}{c_s^2} - \frac{k^2}{a^2}\cR_k^2 \right) + \frac{\cA_\N^2}{2L^4H^2} \( \frac{k^4}{a^4} \)\cR_k^2 \] \,,
\end{align}
with $\tilde\varepsilon$ defined in \pref{tildeeps} and where
\begin{align}\label{ets1}
\frac{1}{c_s^2} := \( \frac{\dot\varphi^2}{2\tilde\varepsilon\mpl^2H^2} \) \frac{\cF^2_{}}{L^4} \,.
\end{align}
In the regime where we derived the previous expressions, the last term in \eqref{csMI} is always subdominant, and we find a standard mode 
with a speed of propagation $c_s^2\sim \sqrt{\tilde \varepsilon} \, m^2/L^2$ which may be subluminal depending on the parameters of the model. 
This model seems to be free from any pathologies.

\subsubsection*{Chromo-natural inflation}

The case of Chromo-natural inflation provides a concrete check on the above discussion. As we discussed in  sec.~\ref{sec:ch}, magnon inflation 
is related to Chromo-natural inflation at the background level. Concerning perturbations, since we work in the case with $U^\mu=u^\m$, one can forget about the order parameter and simply upgrade the potentials in \pref{CNLagII} to the potentials in the theory.
The background equations in this case are well approximated by
\begin{align}
\frac{\dot{\chi}}{\tilde{f}} = \frac{\tilde{g}\psi}{\lambda}, \quad \psi^3 = \frac{\tilde\mu^4\sin\(\frac{F \chi}{\tilde{f}}\)}{3\tilde{g}H\lambda F^3}.
\end{align}
This implies
\begin{align}
  \T_a \, \exd\phi^a  = \exd\chi, \quad \N_a \,\exd\phi^a  = \exd\psi,
\end{align}
while we can evaluate  (cf. \eqref{eq:potch})
\be
\cF = -\frac{3 \tilde g \lambda F^4}{\tilde{f}} \,  \psi^2, \quad \M_{\chi\chi}  = 0  \quad \hbox{and} \quad \M_{\psi\psi} =  6 \tilde g^2 F^2\psi^2 \,.
\ee
Also, note that $\A_\N = 0$, so that our unitary gauge quadratic action, eq.\ \eqref{effact}, is particularly simple,
\be
S^{(2)}  =   \int \frac{\exd^3  k}{(2\pi)^3}\,\exd t\, a^3 \mpl^2  \varepsilon \Bigg[ 3 \dot{\curv}_k^2-  \frac{k^2}{a^2}\curv_k^2  
\Bigg] \,,
\ee
showing how Chromo-natural inflation resembles a model with a reduced sound speed $c^2_s  = 1/3$, as was noted previously in \cite{Dimastrogiovanni:2012st,Adshead:2013nka}.

\section{Conclusion}
\label{section:discussion}

In this paper we study a class of multi-scalar effective field theories (EFTs) that can achieve inflationary slow roll despite having a scalar potential that does not satisfy $\cG^{ab} \partial_a V \partial_b V \ll V^2/\mpl^2$ (where $\cG_{ab}$ is the target-space metric). They evade the usual slow-roll conditions on $V$ because their kinetic energies are dominated by single-derivative terms rather than the usual two-derivative terms. The presence of such terms requires some sort of UV Lorentz-symmetry breaking during inflation (besides the usual cosmological breaking, although at low enough energies their implied preferred frames naturally align). Chromo-natural inflation provides an example of a UV theory that can generate the multi-field single-derivative terms we consider and we argue that the EFT we find indeed captures the slow-roll conditions for the background evolution for  Chromo-natural inflation. Truncated to a single field, our EFT superficially resembles Cuscuton-like models at low energies (where the $U^\mu$ appear as auxiliary fields and can be integrated out). The multi-field case introduces a new feature however: the scalar kinetic terms define a target-space 2-form, $\cF_{ab}$, whose antisymmetry gives new ways for slow roll to be achieved.

We find examples within this class of EFTs that can, but need not, cross the phantom divide by giving $w = p/\rho < -1$. This raises the possibility of unstable fluctuations. A preliminary examination indicates that stability of the $w<-1$ regime in general depends on the details of the model, and need not imply instability. However, in some instances (such as when $\cA_\cN = 0$) $w < -1$ does  lead to unstable modes once coupled to gravity.  The case  with $w > -1$ can be easily made stable for the modes described by our EFT.

We remark in closing that although it may seem tempting to consider applying this EFT to model dark energy rather than inflation (see  \cite{Blas:2011en} for a single field example), one would then be forced to confront  strong observational constraints on Lorentz breaking during the present cosmological epoch \cite{Blas:2014aca}.

\section*{Acknowledgements}

We thank Azadeh Maleknejad for the conversation that initiated this line of thought, as well as Gregory Gabadadze, Louis Leblond and Sarah Shandera for useful discussions about inflationary models with preferred frames, and Niayesh Afshordi, Ghazal Geshnizjani and Sergey Sibiryakov for comments on the draft. The Aspen Center for Physics, the Kavli Institute for Theoretical Physics (KITP) and the NYU Center for Cosmology and Particle Physics (CCPP) kindly supported and hosted various combinations of us while part of this work was done. This research was supported in part by funds from the Natural Sciences and Engineering Research Council (NSERC) of Canada and funds from the Swiss National Science Foundation. Research at the Perimeter Institute is supported in part by the Government of Canada through Industry Canada, and by the Province of Ontario through the Ministry of Research and Information (MRI). Work at KITP was supported in part by the National Science Foundation under Grant No. NSF PHY11-25915. Work at Aspen was supported in part by National Science Foundation Grant No. PHYS-1066293 and the hospitality of the Aspen Center for Physics.

\appendix

\section{Calculation of the stress tensor}
\label{app:DeltaTcalc}

We compute here the stress tensor of the two-derivative terms. Starting with $\Delta \cL = \sqrt{-g} \; \Delta L$ we have 
\be
 \delta \Delta \cL = \sqrt{-g} \left[ \frac12 \, \Delta L \, g^{\mu\nu} \, \delta g_{\mu\nu} + \delta  \Delta L \right] \,,
\ee
where we write $\Delta L = \Delta L_{\rm kin} + \Delta L_{\EA} + \Delta L_{56}$, with
\bea
 \Delta L_{\rm kin} &:=& - \frac12 \left( \cG_{ab} \, g^{\mu\nu} + \cI_{ab} \, U^\mu U^\nu \right) \partial_\mu \phi^a \, \partial_\nu \phi^b \,, \nn\\
 \Delta L_{56} &:=& - \cC^{(5)}_a \, \dot U^\mu \partial_\mu \phi^a - \cC^{(6)}_a \, (\nabla \cdot U) \dot \phi^a \,, \\
\hbox{and} \quad
 \Delta L_{\EA} &:=& - \frac12 \, {K^{\mu\nu}}_{\alpha \beta} \, \nabla_\mu U^\alpha \, \nabla_\nu U^\beta \,, \nn
\eea
with dot denoting $U^\mu \nabla_\mu$ as in the main text and
\be
 {K^{\mu\nu}}_{\alpha\beta} := \cC^{(1)} \, g_{\alpha \beta} \, g^{\mu\nu} + \cC^{(2)} \, \delta^\mu_\alpha \, \delta^\nu_\beta + \cC^{(3)} \, \delta^\mu_\beta \, \delta^\nu_\alpha + \cC^{(4)} \, g_{\alpha \beta} U^\mu U^\nu \,.
\ee

The required metric variation is
\be
 \delta  \Delta L = - \frac12 \, \cG_{ab} \, \partial_\mu \phi^a \, \partial_\nu \phi^b \, \delta g^{\mu\nu} - \frac12 \, \delta {K^{\mu\nu}}_{\alpha\beta} \, \nabla_\mu U^\alpha \, \nabla_\nu U^\beta - {J^{\mu}}_{\alpha} \, \delta \left( \nabla_\mu U^\alpha \right)\,,
\ee
where 
\be
 {J^\mu}_\alpha = {K^{\mu\nu}}_{\alpha \beta} \nabla_\nu U^\beta + \cC^{(5)}_a \, U^\mu\partial_\alpha \phi^a + \cC^{(6)}_a \, \dot \phi^a \, \delta^\mu_\alpha \,,
\ee
and 
\be
 {J^\mu}_\alpha \, \delta (\nabla_\mu U^\alpha ) = {J^\mu}_\alpha \, U^\lambda \, \delta \Gamma^\alpha_{\mu\lambda} =  \frac12 \, J^{\mu\rho} \, U^\lambda \, \left( \nabla_\mu \delta g_{\lambda\rho} + \nabla_\lambda \delta g_{\mu\rho} - \nabla_\rho \delta g_{\mu \lambda} \right) \,.
\ee
After an integration by parts, 
\bea
 \delta \Delta L &=& \frac12 \left[\cG_{ab} \, \partial^\mu \phi^a \, \partial^\nu \phi^b + \cC^{(1)} \, \left( \nabla^\mu U^\lambda \nabla^\nu U_\lambda - \nabla_\lambda U^\mu \nabla^\lambda U^\nu \right) - \cC^{(4)} \dot U^\mu \dot U^\nu \right] \delta g_{\mu\nu} \nn\\
 && \qquad\qquad + \frac12 \, \nabla_\lambda \left( J^{\lambda \nu} U^\mu +  J^{\mu\nu} U^\lambda - J^{\mu\lambda} U^\nu \right) \delta g_{\mu\nu}  \,.
\eea

We are led in this way to the following stress-energy contribution,
\bea
 \Delta T^{\mu\nu} &:=&  \frac{2}{\sqrt{-g}} \, \left( \frac{\delta \Delta S}{\delta g_{\mu\nu}} \right)  \\
 &=& \Delta L \, g^{\mu\nu}  + \cG_{ab} \, \partial^\mu \phi^a \, \partial^\nu \phi^b + \cC^{(1)}  \left( \nabla^\mu U^\lambda \nabla^\nu U_\lambda - \nabla_\lambda U^\mu \nabla^\lambda U^\nu \right) - \cC^{(4)} \dot U^\mu \dot U^\nu \nn\\
 && \qquad + \frac12 \nabla_\lambda \left[ \left( J^{\lambda \nu} - J^{\nu\lambda} \right) U^{\mu} + \left( J^{\lambda \mu} - J^{\mu\lambda} \right) U^{\nu} + \Bigl( J^{\mu\nu} +  J^{\nu\mu} \Bigr) U^\lambda \right] \,. \nn
\eea

\section{Perturbations in magnon inflation}
\label{app:MIPerts}

In this section we sketch the derivation of the quadratic action in various gauges for cosmological perturbations in magnon inflation. For simplicity, we restrict ourselves to models with two scalar fields, however, generalizations to higher dimensions is straightforward. We begin by carefully parameterizing the time-dependent background field trajectories, and the spacetime-dependent perturbations about them. We then compute the quadratic fluctuation action in comoving gauge. 

\subsection{Parameterization of the background trajectories}

We first recast some facts about the background. In a general multi-field context, one gains geometrical intuition of the nature of the adiabatic and entropy modes in different gauges by going to the Frenet-Serret formalism \cite{GrootNibbelink:2000vx, GrootNibbelink:2001qt, Achucarro:2010da, Achucarro:2012sm, Burgess:2012dz}. That is, for a given background solution $\varphi^a(t)$, we can construct the tangent vector along the trajectory\footnote{In what follows, we restrict ourselves to a flat two dimensional target space. One can generalize this straightforwardly \cite{GrootNibbelink:2000vx, GrootNibbelink:2001qt, Achucarro:2010da, Achucarro:2012sm}.} 
\begin{align}\label{tangent}
\T^a := \frac{\dot\varphi^a}{\dot\varphi},\,\quad \dot\varphi := \left(\dot\varphi^a \dot\varphi_{a}\right)^{1/2},
\end{align}
where indices are raised and lowered with the flat target space metric $\delta^a_b$. Along with the corresponding normal vector to the trajectory
\begin{align}
\N^a: = \delta^{ac}\epsilon_{cb}\T^b,
\end{align}
with $\epsilon_{ab}$ the antisymmetric pseudo-tensor. Together, $\T^a$ and $\N^a$ are a complete field space basis in two dimensions. The time derivatives of these satisfy the Frenet-Serret relations:
\begin{align}
\dot \T^a = -\dot\vartheta \N^a,\quad \dot \N^a = \dot\vartheta \T^a,
\end{align}
which follow from their normalization and orthogonality, and $\dot\vartheta$ corresponds to an angular velocity in field space. From the background equation of motion, eq.\ \eqref{Seom1}, 
the anti-symmetry of $\widetilde \cF$ implies that the quantity in the square brackets is orthogonal to $\dot\varphi^a$ and therefore to $T^a$, hence
\begin{align}
\N^a = \frac{n^a}{\sqrt{n^an_a}}~,\quad n_a =  3H \cA_b - \partial_b \cV
\end{align}
Furthermore, since $\widetilde \cF^{ab} = -\epsilon^{ab}/\cF$ ($\cF := \cF_{\N\T} = \frac12 \epsilon^{ab} \cF_{ab}$ is the target-space field strength) we find that 
\begin{align}
\dot\varphi^a\dot\varphi_a = \cF^{-2}\,n^an_a.
\end{align}
From eq.\ (\ref{tangent}) we see that
\begin{align}
\dot \T^a = \frac{\ddot \varphi^a}{\dot\varphi}- \frac{\dot\varphi^a}{\dot\varphi^{3}}\ddot\varphi_c\dot\varphi^c,
\end{align}
so that given that $\dot \T^a \N_a = -\dot\vartheta$, one evaluates
\begin{align}
\dot\vartheta = \cF^{-1}\left[ 3 H\cA_{\T,\T} +3 \cA_\T\frac{\dot H}{\dot\varphi} -  \cV_{\T\T} \right],  
\end{align}
where the overdot is shorthand for $U^\mu\nabla_\mu$, and $\cV_{\T\T} := \T^a \T^b\cV_{,ab} $. This is to be compared with the usual expression for a two derivative kinetic term coupled to a potential, where $\dot\vartheta = \cV_\N/\dot\varphi$ \cite{Achucarro:2012sm}. This highlights the novel aspects of the dynamics of magnon inflation -- the two form field strength plays a privileged role in determining the acceleration of the trajectory.

\subsection{Perturbations about the background}\label{sec:app_pert}

We can now address perturbations. Without loss of generality, one can write an arbitrary field profile as \cite{Achucarro:2012sm}
\eq{fsexp}{\phi^a({\bf x}, t) = \varphi^a(t + \pi({\bf x}, t)) + \N^a(t + \pi({\bf x}, t))\sigma({\bf x}, t).}
That is, any field perturbation can be parametrized as a local rescaling of the background solution $\varphi^a(t)$ (the adiabatic mode) plus that part which is left over, necessarily orthogonal to the background trajectory at the rescaled time (the isocurvature mode) \cite{Achucarro:2012sm, Gordon:2000hv}. One proceeds by parametrizing the metric perturbations using the ADM decomposition 
\eq{adm}{ds^2 = -N^2dt^2 + h_{ij}(dx^i + N^idt)(dx^j + N^jdt),}
where different gauge choices correspond to different foliations, and consequently different choices for the induced metric $h_{ij}$.  

The action we are to perturb can be separated into the Einstein-Hilbert term plus the matter action. The matter sector action is given by 
\be
\label{act2}
 S_\ssM =  \int \exd^4x \sqrt{h}N \; \left[ -\cV(\phi) -  \cA_a(\phi) U^\mu \partial_\mu \phi^a - \xi \, \Bigl( g_{\mu\nu} U^\mu U^\nu + 1 \Bigr) \right].
\ee
In the ADM decomposition \cite{Arnowitt:1962hi}, the Einstein-Hilbert action reads
\begin{align}\label{eqn:g}
S_G = \frac{\mpl^2}{2}\int d^4x \sqrt{h}\left[N R^{(3)} + \frac{1}{N}(E^{ij}E_{ij} - E^2)\right],\quad E_{ij} = N K_{ij} = \frac{1}{2}\left[\dot h_{ij} - \nabla_iN_j - \nabla_jN_i\right],
\end{align}
where $R^{(3)}$ is the Ricci scalar constructed out of $h_{ij}$,
and  $K_{ij}$ is the extrinsic curvature of the foliation defined by our gauge choice. We perturb the lapse and shift as
 \begin{align}
N = & 1 + \alpha, \quad \alpha = \alpha_1+\alpha_2+\ldots,\\
N^i = & h^{ij}\partial_j\theta + N^{T\,i}, \quad \theta = \theta_1+\theta_2+\ldots.
\end{align}
As is well-known, we only need the solutions for $\alpha$ and $\theta$ to linear order to find the action up to cubic order \cite{Maldacena:2002vr}. We ignore vector and tensor fluctuations of the metric, and restrict our attention to the scalar sector.

\subsection*{Perturbations in unitary  (or comoving) gauge}
Unitary gauge is defined by setting $\pi \equiv 0$ in eq.\ (\ref{fsexp}), such that this scalar fluctuation is absorbed by the metric 
\begin{align}
h_{ij} \equiv a^2(t)e^{2\cR}\delta_{ij},
\end{align}
and the only field fluctuations are orthogonal to the inflationary trajectory
\begin{align}
\phi^a({\bf x},t) =  & \varphi^a(t) + \N^a(t)\,\sigma({\bf x}, t).
\end{align}
In this gauge, to quadratic order the gravitational part of the action can be written\footnote{We use the notation $\partial^2 = \delta^{ij}\partial_i \partial_j$, and e.g. $(\partial\curv)^2 = \delta^{ij}\partial_i \curv\partial_j \curv$}
\begin{align}
S^{(2)}_{G}=    \frac{\mpl^2}{2}\int \exd^4 x  \Bigg[& -2a\big[ 2\partial^{2}\curv \alpha-(\partial\curv)^{2}\big]  +4a((-\alpha H+\dot{\curv})\partial^2\theta) \\ \nn&  -6a^3 \(H^2\alpha^2- 3H^2\alpha \curv- 2H\dot\curv\alpha+3a^{-3}H\partial_t(a^3 \curv^2)+\dot\curv^{2}-\frac{9}{2}H^2 \curv^2\)\Bigg].
\end{align}
We perturb the Lagrange multiplier field, $\xi$, as well as the contra-variant vector $U^{\mu}$ as
\begin{align}
\xi = & \bar\xi+\delta_1\xi +\delta_2\xi, \quad 
U^{\mu} =  \overline{U}^{\mu}+\delta_1 U^{\mu} + \delta_2 U^{\mu},
\end{align}
again, we only need the linear order field fluctuations, since at quadratic order the second order perturbations simply multiply the background equations of motion and constraints.

To quadratic order, the matter sector action at eq.\ \eqref{act1} becomes
\begin{align}
\mathcal{S}^{(2)}_m 
= & \int \exd^3x \, \exd t \; a^3\Bigg[ -\(1+3\curv+\alpha+3\curv\alpha +\frac{9}{2}\curv^2\){\V}- (1+\alpha+3\curv)\V_{,\N }\sigma -\V_{,\N\N} \frac{\sigma^2}{2}\\\nn
&-\(1+3\curv +\alpha+ 3\curv \alpha + \frac{9\curv^2}{2}\)\dot{\varphi}{\A}_\T-\(1+3\curv +\alpha\)\(\dot{\varphi}\A_{\T,\,\N}\sigma+ {\A}_a\frac{d}{dt}(\N^a\sigma)+\dot{\varphi}{\A}_\T \delta_1U^{0}\)\\\nn
 & -{\A}_a \delta_1U^{0}\frac{d}{dt}(\N^a\sigma)-\dot{\varphi}\A_{\T,\,\N}\sigma \delta_1U^{0} -\A_{a,\,\N}\sigma \frac{d}{dt}(\N^a\sigma) - \frac{1}{2}\dot{\varphi}\A_{\T,\, \N\N}\sigma^2 \\\nn &  -{\A}_\N \delta_1U^{i} \partial_i\sigma + \bar\xi \big((2\alpha+6\curv\alpha+3\alpha^2)+2(1+3\curv+3\alpha)\delta_1 U^{0} +\delta_1 U^{0}\delta_1 U^{0}\big) \\
& \nn +2\delta_1 \xi (\alpha  +  \delta_1 U^{0} ) -\bar\xi h_{ij}(N^i + \delta_1 U^i)(N^j + \delta_1 U^j)
\Bigg] \,.
\end{align}
Varying the action with respect to $\delta_1 \xi$, again yields the constraint
$
\delta_1 U^{0} = -\alpha.
$ 
Further, the field redefinition $\delta_1\tilde{U}^i = N^i + \delta_1 U^i$ removes the quadratic term in $N^i$, so that it remains a Lagrange multiplier field when combined with the gravitational action. Substituting this into the action, dropping constant and linear terms and making use of the background equations of motion, we find
\begin{align}\nn
S^{(2)}_m =  \int \exd^3x \, \exd t \; a^3\Bigg[& - \(3\curv\alpha +\frac{9}{2}\curv^2\){\V}- (\alpha+3\curv)\V_{,\N }\sigma -\V_{,\N\N} \frac{\sigma^2}{2}- \frac{9\curv^2}{2}\dot{\varphi}{\A}_\T \\ & -3\curv \(\dot{\varphi}\A_{\T,\,\N}\sigma+ {\A}_a \frac{d}{dt}(\N^a\sigma)\)-\A_{a,\,\N}\sigma U^{0}\frac{d}{dt}(\N^a\sigma) \\\nn & - \frac{1}{2}\dot{\varphi}\A_{\T,\, \N\N}\sigma^2   -{\A}_\N (\delta_1\tilde{U}^j - N^j)\partial_j\sigma   
-a^2 \bar\xi \delta_{ij}\delta_1 \tilde{U}^i \delta_1 \tilde{U}^j
\Bigg]
\end{align}
combining with the gravitational action, and using the Friedmann equation, we have
\begin{align}
S^{(2)}=  \int \exd^3 x \, \exd t \; a^3 \Bigg[ &-a^{-2}\mpl^2\big[ 2\partial^{2}\curv \alpha-(\partial\curv)^{2}\big]  +2a^{-2}\mpl^2 (-\alpha H+\dot{\curv})\partial^2\theta \\ \nn&  -3 \mpl^2 \(H^2\alpha^2- 2H\dot\curv\alpha+3a^{-3}H\partial_t(a^3 \curv^2)+\dot\curv^{2}\)- (\alpha+3\curv)\V_{,\N }\sigma  \\ \nn& -\V_{,\N\N} \frac{\sigma^2}{2}- \frac{9\curv^2}{2}\dot{\varphi}{\A}_\T  
-3\curv \(\dot{\varphi}\A_{\T,\,\N}\sigma+ {\A}_a \frac{d}{dt}(\N^a\sigma)\)-\A_{a,\,\N}\sigma \frac{d}{dt}(\N^a\sigma) \\ \nn& - \frac{1}{2}\dot{\varphi}\A_{\T,\, \N\N}\sigma^2    -{\A}_\N (\delta_1\tilde{U}^j - \partial^j\theta)\partial_j\sigma   
- a^2 \bar\xi \delta_{ij}\delta_1 \tilde{U}^i \delta_1 \tilde{U}^j
\Bigg]
\end{align}
Variation of the action with respect to $\theta$ yields an equation for the perturbation to the lapse,
\begin{align}
\alpha   = \frac{\dot{\curv} }{H} -  \frac{{\A}_\N  }{2 H \mpl^2 } \sigma.
\end{align}
Variation with respect to $\alpha$ yields an equation for $\theta$, however, we do not need it at this order in perturbation theory. This is because it appears only linearly in the action, and thus simply multiplies its own equation of motion.

Substituting in, after much algebra and making use of the background equations of motion we arrive at
\begin{align}
S^{(2)}    =   \int \exd^3  x\, \exd t \; a^3 \Bigg[ & -a^{-2}\mpl^2 \varepsilon \(\partial\curv\)^2   + a^{-2}\frac{\A_\N}{H }  \partial^{2}\curv \sigma  +\cF_{\N\T} \dot{\varphi} \sigma\frac{\dot\curv}{H}   \\\nn
 & 
-\(\M_{\N\N}-\M_{\T\T}+\frac{\A_\N }{2 H \mpl^2 }\dot{\varphi}  \cF_{\N\T}\)\frac{\sigma^2}{2}-  {\A}_\N \delta_1\tilde{U}^{j} \partial_j\sigma  - \frac{\dot{\varphi}{\A}_\T}{2}\(a^2\delta_{ij}\delta_1 \tilde{U}^i \delta_1 \tilde{U}^j \)  
\Bigg]
\end{align}
where $\M_{ab}$ was defined above as
\begin{align}
\M_{ab}  := \F_{ad}\[\tilde{\F}^{dc}(\V_{,c}-3H\A_c)\]_{,b}.
\end{align}
Note that, if $\dot{\varphi}{\A}_\T \neq 0$, which is equivalent to $\varepsilon\neq 0$, then we can integrate out $\delta_1 U_i$ to find
\begin{align}
S^{(2)}    =   \int \exd^3  x\, \exd t \; a^3 \Bigg[& -\frac{\mpl^2 \varepsilon}{a^2} \(\partial\curv\)^2   + \frac{\A_\N}{a^2 H }  \partial^{2}\curv \sigma  +\cF_{\N\T}  \dot{\varphi}\sigma\frac{\dot\curv}{H}  \\\nn
 &     
-\(\M_{\N\N}-\M_{\T\T}+ \frac{\A_\N }{2 H \mpl^2 }\dot{\varphi}  \cF_{\N\T}\)\frac{\sigma^2}{2}+ \frac{{\A}_\N{\A}_\N}{2a^2\dot{\varphi}{\A}_\T}\(\partial_j\sigma\partial_j\sigma\)  
\Bigg]
\end{align}
We observe that the nominal isocurvature mode $\sigma$ is an auxiliary field so we are entitled to integrate it out. The quadratic Lagrangian has the form
\begin{align}\label{pert1}
\cL^{(2)} = -\frac{L^4}{2}\sigma^2 +\frac{\cA_\N}{a^2H}\sigma\partial^2\calR - \mpl^2\frac{\varepsilon}{a^2}(\partial\calR)^2 + \frac{\cA_\N^2}{2a^2\dot\varphi\cA_\T}(\partial\sigma)^2  +\cF_{\N\T}\frac{\dot\varphi}{H}\,\sigma\dot\cR 
\end{align}
with
\begin{align}
  L^4 =  \(\M_{\N\N}-\M_{\T\T}+\frac{\A_\N }{2 H \mpl^2 }\dot{\varphi} \cF_{\N\T}\).
\end{align}
Formally integrating out $\sigma$, one obtains the Lagrangian
\eq{}{\cL^{(2)} = -\mpl^2\frac{\varepsilon}{a^2}(\partial\calR)^2 +\frac{1}{2}\left(\frac{\dot\varphi}{H} \cF_{\N\T}\dot\cR + \frac{\cA_\N}{a^2H}\partial^2\cR \right)\widehat\Omega^{-1} \left(\frac{\dot\varphi}{H} \cF_{\N\T} \dot\cR + \frac{\cA_\N}{a^2H}\partial^2\cR \right) }
where
\begin{align}
\widehat\Omega := -\frac{\cA_\N^2}{2H^2 \mpl^2 \varepsilon}\frac{\partial^2}{a^2} + L^4,
\end{align}
having used eq.\ \eqref{alt-eps}. Fourier transforming and integrating the cross term, $\dot{\cR}\partial^2\cR$, by parts and discarding the boundary terms gives the result in eq.\ \eqref{effact}.

  \bibliographystyle{JHEP}
  \bibliography{FMI}

\end{document}